\documentstyle[aps,prc,epsf]{revtex}

\begin{document}
\small
\newcommand {\qbar} {\=q}
\newcommand {\sbar} {\=s}

\draft
\title{Quark Molecular Model of the $S=0$
Strange Pentaquark (u\sbar)-(uds) Baryon Spectrum}
\author{Robert A. Williams}
\address{Hampton Roads Academy, Newport News, VA 23608}
\author{Paul Gu\`eye}
\address{Physics Department,Hampton University, Hampton, VA 23668 }
\date{\underline{Draft:} \today}

\maketitle
\begin{abstract}
We present a non-relativistic quark molecular model (QMM) of the strange crypto-exotic pentaquark
baryon spectrum motivated by the recent data showing narrow, resonance enhancements in
electromagnetic and hadronic production of kaons. Our model assumes color octet bonded quark
molecular clusters forming exotic color singlet pentaquark baryons. We develop explicit molecular
pentaquark wavefunctions that exhibit color interchange and cluster fermion antisymmetry. 
Our color electro-dynamics (CED) inspired Hamiltonian, which includes confinement, one gluon
exchange and color magnetic interactions, is a natural generalization of the Isgur and Karl  
quark model Hamiltonian that reproduces the conventional meson (q\qbar) and baryon (qqq)
spectrum~\cite{Isgur77}. We introduce the idea of Color Magnetic Confinement (CMC) which is
employed to restrict the physical pentaquark spectrum and constrain the fundamental constants 
(color gyromagnetic ratios) of the QMM-CED Hamiltonian. Comparing our QMM spectrum predictions
with the recent experimental data in associated strangeness photo- and electro-production reactions,
we identify 4 candidate states that have masses and total widths consistent with exotic pentaquark
structure.
\end{abstract}

\pacs{PACS number(s): 12.39.Jh; 24.30.Gd; 13.40.-f; 13.75.-n}

\section{Introduction}

Recent experimental data on electromagnetic and hadronic associated strangeness production suggest
the possible existence of narrow multi-quark $S = 0$~\cite{GW01,GW02} and $S=+1$~\cite{SAPHIR03,CLAS03}
states near 2.0 GeV. These pentaquark systems are in the realm of opening a complete new area of
exciting physics on both experimental and theoretical fronts.

The existence and corresponding properties of exotic matter remains a puzzling open question in
hadron spectroscopy. Resolving this mystery is important for understanding how QCD provides a
consistent description for the binding of quarks and gluons into hadrons~\cite{Lipkin87}. Exotic
hadrons are color singlet multiquark combinations (qq\qbar\qbar, qqqq\qbar ~\ldots) beyond the
standard valance  q\qbar-(meson) and qqq-(baryon) quark content. Crypto-exotics are those multiquark 
states which have degenerate quantum numbers with the normal hadrons (i.e., {\it non-exotic}
quantum numbers). Examples include the $a_0/f_0(980)$ resonances, which are established scalar
meson states that find their most natural interpretation as hadronic molecules, having wave
functions with large components of lightly bound K\=K mesons~\cite{WeinIsgur90,Schumacher97}.
The most stable configurations of qqqq\qbar ~exotics are believed to be strange molecules, like
$S=0$ (uds)-(u\sbar) or $S=+1$ (udd)-(d\sbar), bound by long range color forces acting between color
octet quark clusters (octet-bonded states). Exotic molecules such as (qq)-(qq\qbar) can exist in
the color 6-\=6 representations (sextet-bonded states), however these are expected to be much less
stable because a color triplet qq\qbar ~subsystem always contains a q\qbar ~pair which can form a
color singlet, allowing the sextet bonded qqqq\qbar ~system to easily dissociate into a normal
color singlet baryon (qqq) and meson (q\qbar)~\cite{CHS79,HS78}. For a baryon $B \equiv qqq$ and
a meson $M \equiv$ q\qbar, if $B$ and $M$ are separately color singlets, the molecules $(B-M)$ will
interact only via the relatively weak  Van der Waals forces of QCD and thus will dissociate
rapidly (and have large decay width) into a normal baryon and meson. However, if the molecular
system $(B-M)$ is a color singlet with $B$ and $M$ being separately color octet states and
possessing non-zero relative orbital angular momentum $L$, this system should be linked by very
strong color forces and exhibit enhanced stability (i.e., long lifetime and narrow decay width 
compared to usual hadronic resonance states) since the anti-quark in $M$ is prevented by a
centrifugal barrier from tunneling into $B$ where it can combine with a quark to form a color
singlet meson. Additional stability is anticipated if at least one molecule cluster contains a
strange (or charm) quark, since then decays to open non-strange ($\pi, \rho, \omega$) channels
will be OZI~\cite{OZI} suppressed.

For $S=0$ strange nucleon resonances, each color octet molecule cluster must have a strange
quark, $B = qqs$ and $M = q\bar{s}$. This naively paradoxical feature  of having $S=0$ with
strange exotic structure generates an interesting selection rule for detecting exotic baryons:
the most stable pentaquark baryons will preferentially decay into strange meson and baryon
channels by the OZI rule (such as $K^+ \Lambda$, $K^{*+} \Lambda$,  $K^+ \Sigma$, etc.), 
since the strangeness in each molecular $B$ and $M$ cluster is conserved in the OZI enhanced  
final states.  Consequently, the $S=0$ pentaquark exotics will potentially show up as narrow
nucleon resonances in kaon ($K$ or $K^*$) photo- and electro-production reactions, yet totally
decouple from the non-strange $\pi,\rho,\omega$ production channels (or possibly make very small
OZI evading/violating contributions).

This paper is the first work to suggest the possibility of exciting $S=0$ pentaquark resonances
in single kaon electromagnetic production from the nucleon (proton or neutron). The recently
reported $S=+1$ pentaquark baryon resonance in $K^+ K^-$ photo-production from the
neutron~\cite{SAPHIR03}  is a crypto-exotic state with $\Lambda$ quantum numbers (similar to the
$\Lambda(1405)$ which is widely accepted to be a $K^-p$ molecule). In this work we restrict our
analysis to the $S=0$ nucleon resonance sector, leaving  the $S=\pm 1$ exotics for consideration
in a follow up paper. Before passing into the formalism section, we note that previous authors 
claim that the $S=0$ strange exotic baryons and their excited states are expected to fall on
linear Regge trajectories~\cite{CJKTW74,JT76}.

The excited molecular pentaquarks  presumably also have anomalously narrow decay widths in
comparison to {\it normal}  excited baryons since the same enhanced stability arguments apply.  
In contrast, our Quark Molecular Model (QMM) implementation permits only 4-low lying $S=0$
pentaquark solutions, and dynamically excludes the  possibility of additional excited states.
This limited spectrum feature is a novel prediction of  this QMM and stands in contrast to the
usual outcome of quark model spectroscopy: to generate an  overabundance of excited states, and
providing theoretical bias for the experimental dilemma  of ``{\it missing resonances}''.

This paper focuses on two main areas: (1) a dynamical quark model of the $S=0$ strange exotic
baryon (u\sbar)-(uds) spectrum; and (2) new/recent experimental data that provides preliminary
(albeit inconclusive) support for the possible existence of strange exotic resonances.

Section~\ref{Formalism} describes the general formalism of our quark molecular model (QMM), and
section~\ref{ExperimentalStatus} describes the current status of experimental data that suggest
resonance enhancements due to the $S=0$ pentaquark baryons. In forthcoming papers we will present
a generalized hadrodynamical Lagrangian model that incorporates exotic resonances through strong
final state interactions in kaon photo- and electro-production reactions~\cite{GW01,GW02}, where
we will demonstrate that the exotics have a strong forward angle production sensitivity.

\section{Formalism}\label{Formalism}

Our approach is to first construct totally antisymmetric spin-flavor-space-color pentaquark baryon 
wavefunctions, and then calculate the low lying crypto-exotic baryon spectrum using a non
relativistic quark molecular model (QMM) basis for the spatial wavefunctions and energy eigenvalues. 
In this work, we assume the pentaquark crypto-exotic systems are heavy, color octet bonded $(B-M)$
molecules with anomalous stability. Dynamical stability and explicit decay width calculations 
will be investigated in a future work. Based on the ideas of our introductory discussion, the
lightest, stable $S=0$ exotic baryons are expected to have two distinct features: (1) contain
strange quark molecule clusters $B = qqs$ and $M = q\bar{s}$ to be OZI protected from non-strange
decay channels~\cite{OZI}; (2) molecule clusters possess at least one unit of relative orbital
angular momentum to generate a centrifugal barrier which suppresses quark rearrangement decay processes.
In the derivation that follows, we show that both of these qualitative features naturally arise
from an explicit minimal construction of pentaquark wavefunctions with color octet bonded molecular 
substructure when total antisymmetry (Fermi statistics) of quark interchange is enforced.   

In order to construct a minimal color singlet pentaquark molecule with total fermion antisymmetry, 
first consider the normal 3-q baryon $SU(3)_F$ flavor octet and decuplet wavefunctions. Normal
ground state 3-q baryons have totally antisymmetric quark combinations which are products of color 
[antisymmetric (A) singlet: $C_A^1$], flavor [mixed symmetry (MS) octet: $F_{MS}^8$; or symmetric 
(S) decuplet: $F_{S}^{10}$], spin [mixed symmetry (MS) spin $\frac{1}{2}$ doublet: 
$\chi_{MS}^{\frac{1}{2}}$; or symmetric (S) spin $\frac{3}{2}$ quartet: $\chi_{S}^{\frac{3}{2}} $],
and spatial [symmetric (S) even parity: $\Phi_S^+$] wavefunctions, 
$\Psi_{symmetry}(color,flavor|spin^{parity})$:
\begin{eqnarray}
\Psi_A (1,8|\frac{1}{2}^+) &=& C_A^1 \; \left(F_{MS}^8 \;
	\chi_{MS}^{\frac{1}{2}}\right)_S \Phi_{S}^+ \;\;\; (octet) \\
\Psi_A (1,10|\frac{3}{2}^+) &=& C_A^1 \; F_{S}^{10} \;
	\chi_{S}^{\frac{3}{2}} \; \Phi_{S}^+ \;\;\; (decouplet), 
\end{eqnarray} 
where the large round brackets denotes the symmetric product of $MS$ flavor and spin functions. The
corresponding antisymmetric $(F \chi)_A$ product is excluded for ground states ($L=0$, and even parity)
because the color wavefunction is totally antisymmetric under quark interchange, thus the ground state 
3-q baryons always have an overall symmetric product of flavor-spin-space wavefunctions. It is
interesting to note that there is no flavor singlet (antisymmetric u,d,s wavefunction) 3-q state in
the observed low lying baryon spectrum (the $\Lambda(1405)$ is not a candidate since it is interpreted
as a $K^- p$ molecule, and the ground state $\Lambda$ is the isospin singlet member of the flavor octet).

In the quark model, it is easy to see why there is no flavor singlet baryon (at least, in the
$\frac{1}{2}^+$ or $\frac{3}{2}^+$ ground states). Constructing the quark model as before, and
requiring that all physical hadrons must exist in color singlet configurations, we obtain overall $MS$
and $S$ symmetry for the $\frac{1}{2}^+$ and $\frac{3}{2}^+$ color singlet $\otimes$ flavor singlets: 
\begin{eqnarray}
\Psi_{MS}(1,1|\frac{1}{2}^+) &=& C_{A}^{1} \; F_{A}^{1} \; \chi_{MS}^{\frac{1}{2}} \; \Phi_{S}^+ \\
\Psi_{S}(1,1|\frac{3}{2}^+) &=& C_{A}^{1} \; F_{A}^{1} \; \chi_{S}^{\frac{3}{2}} \; \Phi_{S}^+ .
\end{eqnarray}
Hence ground state flavor singlet baryons are inconsistent with the total anti-symmetry requirement.
However, if the flavor singlet $F_{A}^1(uds)$ $\frac{1}{2}^+$ and $\frac{3}{2}^+$ baryons have octet
color symmetry, then they can bind with q\qbar\, color octets to form physical color singlet baryon
states. Our QMM is built on the assumption of color octet bonded $\alpha$-$\beta$ molecules between
baryon number $B_{\alpha}=0$ $\alpha$(u \sbar) and $B_{\beta}=1$ $\beta(uds)$ basis states. As
previously discussed, we restrict our attention in this paper to baryon molecules with quantum numbers
$B=1$, isospin $I=\frac{1}{2}$, and hypercharge $Y=B+S=1$, which are crypto-exotics in the nucleon
resonance spectrum (i.e., same quantum numbers as the $N^*$ states).

	\subsection{Color Octet $\bf \alpha$(u \sbar) and $\bf \beta$(uds) Basis States}

We construct totally antisymmetric, color singlet pentaquark baryons which are bound color octet quark
molecule clusters in the form: $(\alpha_8 - \beta_8)_1$, where $\alpha_8$ and $\beta_8$ are generic
labels referring to the color octet $B=0$ $\alpha$(u\sbar,d\sbar) and $B=1$ $\beta$(uds) clusters
respectively. The central idea required to derive explicit wavefunctions in the $\alpha$-$\beta$
QMM basis is to exploit the $SU(3)$ symmetry of usual quark model color and flavor functions to
generate pentaquark states that are color singlets and satisfy quark interchange antisymmetry. 

Consider a color octet baryon state (which, like colorful quarks, can not exist unbound in isolation)
generated by a $SU(3)$ Color $\leftrightarrow$ Flavor symmetry transformation:
\begin{eqnarray}
C_A^1(r,b,g) &\rightarrow & C_{MS}^8(r,b,g) \;=\; 
F_{MS}^8(u\rightarrow r,d\rightarrow b,s\rightarrow g) \;  \\ 
F_{MS}^8(u,d,s) & \rightarrow & F_{A}^1(u,d,s)\;=\; 
C_A^1(r\rightarrow u,b\rightarrow d,g\rightarrow s) ,
\end{eqnarray}
where $C_{A}^1$ and $F_{MS}^8$ are the standard antisymmetric color singlet and mixed symmetry flavor
octet functions written in the literature describing details of the conventional quark model. The
color-flavor (C/F) transformation simply takes the color singlet function, replacing labels (r,b,g) with 
(u,d,s) to generate the flavor singlet flavor function, and the flavor octet functions, replacing 
labels (u,d,s) with (r,b,g) to generate the color octet functions. The C/F transformation generates
one color-octet, flavor-singlet baryon, $\xi(\frac{1}{2}^+)$:
\begin{eqnarray}
|\xi_{8}|(8,1)|\frac{1}{2}^+> &=& 
F_A^1(u,d,s) \; \left(C_{MS}^8(r,b,g) \;  \chi_{MS}^{\frac{1}{2}} \right)_S \Phi_{S}^+ \; ,
\end{eqnarray}
where the MS color and spin function product is symmetrized, and one color-decuplet, flavor-singlet
baryon, $\xi(\frac{3}{2}^+)$:
\begin{eqnarray}
|\xi_{10}|(10,1)|\frac{3}{2}^+> &=&  
F_A^1(u,d,s) \;  C_{S}^{10}(r,b,g) \; \chi_S^{\frac{3}{2}} \; \Phi_S^+ \;.
\end{eqnarray} 

Our goal is to produce overall antisymmetric pentaquark wavefunctions from color octet bonded
molecular clusters $\alpha$(q \sbar) and $\beta$(uds). We accomplish this by forming the
anti-symmetrized product of $\alpha \otimes \beta$ quark cluster wavefunctions with odd ($L=1$) relative
orbital angular momentum:
\begin{eqnarray}
\Psi^X_{\alpha \beta}(J^+) &=& |\left( \alpha \beta \right)_A||B=1,I=\frac{1}{2},Y=1| J^+ >  \\
&=& |\;\alpha(q \bar{s}|0^-,1^-)|| \; C_{MS}^8(\alpha) \;  F_A^2(\alpha)\; 
 \chi_{A,S}^{0,1}(\alpha) \; \Phi_S^-(\alpha)  > \\
&\otimes &
|\;\beta(uds|\frac{1}{2}^+)\;|| \; C_{MS}^8(\beta) \; F_A^1(\beta) \; \chi_{MS}^{\frac{1}{2}}(\beta)\; 
\Phi_{S}^+(\beta) > \nonumber \\
&\otimes & Y_{L M}^{\alpha \beta}(\theta,\phi)|_{L=odd} \; \nonumber  \\
&=& C^1_{A}(\alpha \beta | X) \; 
F_{S}^{2}(\alpha \beta|X) \; \\
&& \left[ \chi_{_{S}}^{2S_X+1}(\vec{S}_X = \vec{S}_{\alpha}+\vec{S}_{\beta}) \; 
\Phi^-_S(\alpha \beta|X) \; \otimes \;
Y_{L \; M}^-(\alpha \beta|X)|_{L=odd} \right]_S^{J = L + S_X} \;. \nonumber
\end{eqnarray}

Here, we suppress Clebsch-Gordon factors and only express how overall fermion and color antisymmetry 
is generated in the QMM basis. The idea is to simultaneously couple the color octet $\alpha$
and $\beta$ functions to an overall antisymmetric $(\alpha - \beta)_A^1$ color singlet function
(to satisfy the quark confinement requirement). For notation when we consider experimental data
later, we refer to the physical ($\alpha-\beta$) pentaquark molecules as the ``$X$'' states. In
the QMM construction, one can see that these $X$ states must have even parity and the
$B=1, I=\frac{1}{2},Y=1$ quantum numbers of nucleon resonances. The color singlet function
$C^1_{A}(\alpha \beta|X)$ (which couples $8_{\alpha} \otimes 8_{\beta} \rightarrow 1_X$)
is totally antisymmetric under color exchange. We construct explicit antisymmetric pentaquark
color functions in the next subsection.

Note the flavor doublet $F_{S}^2(\alpha \beta|X)$ and spin $\chi^{2S_X+1}_{_{S}}(\alpha \beta|X)$
functions individually have $S$-symmetry, and the total $X$ angular momentum $J=L+S_X$ couples
total spin $S_X = S_{\alpha} + S_{\beta}$ and orbital angular momentum $L$ (with appropriate
Clebsch-Gordon factors). A detailed analysis shows that the quarks in the $\alpha$(q\sbar) cluster
can couple to either spin $S=0,1$, in which case the coupled spin-space functions for $\alpha$
must be in $A$ or $S$ configurations respectively (for $S_{\alpha}=0,1$ spin state orthogonality).
Fermion interchange antisymmetry arises from the product of $A$-symmetry color and $S$-symmetry
flavor, spin, and space $(\alpha \beta|X)$ functions. The $S$-symmetry of the iso-doublet molecular
flavor function $F_S^2(\alpha \beta|X)$ is achieved from the product of two totally anti-symmetric
flavor functions: the iso-doublet $F_A^2(\alpha)$ and flavor singlet $F_A^1(\beta)$. It is
important to note here that interchange between quarks in $|\alpha>$ and quarks in $|\beta>$ only
preserve the overall exact fermion antisymmetry if the $(\alpha - \beta)$ molecule has even orbital
angular momentum $L$, because otherwise (for odd $L$) $\alpha \leftrightarrow \beta$ quark exchange
would generate a $MS$ fermion exchange symmetry component due to the negative phase coming from
the odd parity orbital $(\alpha - \beta)$ spherical harmonic $Y^-_{L M}(\alpha \beta|X)$ function.
Our QMM restriction to odd-$L$ states is essentially a dynamical constraint. Primarily,
even-$L=0,2,4,...$ states could not be formed in a {\it cluster basis} because certain quark 
exchanges (like $s$ from $|\beta>$ exchanged with $u$ or $d$ from $|\alpha>$) would generate
wavefunctions which are not {\it cluster factorized} (products of $|\alpha>$ $\otimes$ $|\beta>$
cluster functions) and hence destroys the chance for defining ``basis states'' with invariant
multiplet quantum numbers. Furthermore, $\alpha \leftrightarrow \beta$ quark exchange would allow
$s$ or $\bar{s}$ tunneling and subsequent annihilation, which obviously destabilizes the pentaquark
bound state. In QMM we prohibit $\alpha \leftrightarrow \beta$ quark exchange by requiring 
the ground state to have one unit of relative orbital angular momentum, and excited states must
take odd-$L$ (to generate a centrifugal barrier to dynamically suppress quark rearrangement decay
and to provide the even parity required for crypto-exotic nucleon quantum numbers). So the ground
state $L=1$, and $L=odd$ excited state QMM wavefunctions have a restricted {\it cluster fermion
antisymmetry} (that is, exact $A$-exchange symmetry for permutations within the $\alpha(u,d,s)$
cluster and/or permutations within the $\beta(q,\bar{s})$ cluster, but $MS$-symmetry for
interchanges between clusters). Exact overall fermion antisymmetry could be generated for
$\alpha \leftrightarrow \beta$ quark exchange in odd-$L$ molecules if mixed color symmetry were
allowed, but color confinement requires all physical states to be $A$-color singlets. From this
more fundamental point of view, our QMM cluster fermion antisymmetry is the dynamical symmetry
enforcing exact color confinement in a basis of factorized quark cluster functions with odd
relative orbital angular momentum (and nucleon, $N^*$ quantum numbers). Finally, we observe that
for fermion antisymmetry within the $\beta(uds)$ cluster, the $\beta$(color-spin) functions must
be symmetrized, as indicated in equation (7), since the flavor singlet carries the overall
antisymmetry in $|\beta(u,d,s)>$. 

	\subsection{Pentaquark Wavefunctions}\label{sec:wavefunctions}

		\subsubsection{Color Octet $\beta$(uds) States}

In this work, we restrict our attention to $B=1$ molecules, which must have the flavor singlet 
$\xi(\frac{1}{2}^+)$ as its baryon constituent.  The color decuplet $\xi(\frac{3}{2}^+)$ cannot
form $B=1$ molecules since there is no possible construction of color decuplet q\qbar states,
which would be necessary to form physical, color singlets. However, we note that the
$\xi(\frac{3}{2}^+)$ and its anti-particle $\bar{\xi}(\frac{3}{2}^+)$ could form color singlet
molecule states with baryon number $B=0$ ($\xi_{10}-\bar{\xi}_{\bar{10}}$ mesonic sextaquark) 
or $B=2$ ($\xi_{10}-\xi_{10}$ di-baryons). We write the explicit color-octet (spin-color
symmetrized) flavor-singlet $\xi(\frac{1}{2}^+)$ baryon wavefunctions labeled by {\it color
isospin} ($I^c$,$I^c_3$) and {\it color hypercharge} ($Y^c$): 
$|\xi_{c}(spin^{parity})|I^c,I^c_3,Y^c>$ 
\begin{eqnarray}
p_c^+ &=& |p_c^+(\frac{1}{2}^+)|\frac{1}{2},\frac{1}{2},1> \;= \; |\xi(8,1)| \frac{1}{2}^+|rrb>  \\
&=& \frac{1}{\sqrt{6}}[uds - usd + sud - sdu + dsu - dus] \otimes \\
&& \frac{1}{3\sqrt{2}} [ (2 r_{\uparrow} r_{\uparrow} b_{\downarrow} 
- r_{\uparrow} r_{\downarrow} b_{\uparrow} - r_{\downarrow} r_{\uparrow} b_{\uparrow} )
+ Perms(1 \leftrightarrow 3, 2 \leftrightarrow 3)] \nonumber \\
\nonumber \\
n_c^0 &=& |n_c^0(\frac{1}{2}^+)|\frac{1}{2},-\frac{1}{2},1> \;= \; |\xi(8,1)| \frac{1}{2}^+|bbr>  \\
&=& \frac{1}{\sqrt{6}}[uds - usd + sud - sdu + dsu - dus] \otimes \\
&& \frac{1}{3\sqrt{2}} [ (2 b_{\uparrow} b_{\uparrow} r_{\downarrow} 
- b_{\uparrow} b_{\downarrow} r_{\uparrow} - b_{\downarrow} b_{\uparrow} r_{\uparrow} )
+ Perms(1 \leftrightarrow 3, 2 \leftrightarrow 3)] \nonumber \\
\nonumber \\
\Sigma^+_c &=& |\Sigma^+_c(\frac{1}{2}^+)|1,1,0> \;= \; |\xi(8,1)| \frac{1}{2}^+|rgr>  \\
&=& \frac{1}{\sqrt{6}}[uds - usd + sud - sdu + dsu - dus] \otimes \\
&& \frac{1}{3\sqrt{2}} [ (2 r_{\uparrow} g_{\uparrow} r_{\downarrow} 
- r_{\uparrow} g_{\downarrow} r_{\uparrow} - r_{\downarrow} g_{\uparrow} r_{\uparrow} )
+ Perms(1 \leftrightarrow 3, 2 \leftrightarrow 3)] \nonumber \\
\nonumber \\
\Sigma^0_c &=&|\Sigma^0_c(\frac{1}{2}^+)|1,0,0> \;= \; |\xi(8,1)| \frac{1}{2}^+|rbg(1)>  \\
&=& \frac{1}{\sqrt{6}}[uds - usd + sud - sdu + dsu - dus] \otimes \\
&& \frac{1}{3\sqrt{2}} [ (2 r_{\uparrow} b_{\uparrow} g_{\downarrow} 
- r_{\uparrow} b_{\downarrow} g_{\uparrow} - r_{\downarrow} b_{\uparrow} g_{\uparrow} )
+ Perms(1 \leftrightarrow 3, 2 \leftrightarrow 3)] \nonumber \\
\nonumber \\
\Sigma^-_c &=&|\Sigma^-_c(\frac{1}{2}^+)|1,-1,0> \;= \; |\xi(8,1)| \frac{1}{2}^+|bgb>  \\
&=& \frac{1}{\sqrt{6}}[uds - usd + sud - sdu + dsu - dus] \otimes \\
&& \frac{1}{3\sqrt{2}} [ (2 b_{\uparrow} g_{\uparrow} b_{\downarrow} 
- b_{\uparrow} g_{\downarrow} b_{\uparrow} - b_{\downarrow} g_{\uparrow} b_{\uparrow} )
+ Perms(1 \leftrightarrow 3, 2 \leftrightarrow 3)] \nonumber 
\nonumber \\
\Xi^0_c &=&|\Xi^0_c(\frac{1}{2}^+)|\frac{1}{2},\frac{1}{2},-1> \;= \; |\xi(8,1)|\frac{1}{2}^+|ggr>  \\
&=& \frac{1}{\sqrt{6}}[uds - usd + sud - sdu + dsu - dus] \otimes \\
&& \frac{1}{3\sqrt{2}} [ (2 g_{\uparrow} g_{\uparrow} r_{\downarrow} 
- g_{\uparrow} g_{\downarrow} r_{\uparrow} - g_{\downarrow} g_{\uparrow} r_{\uparrow} )
+ Perms(1 \leftrightarrow 3, 2 \leftrightarrow 3)] \nonumber \\
\nonumber \\
\Xi^-_c &=&|\Xi^-_c(\frac{1}{2}^+)|\frac{1}{2},-\frac{1}{2},-1> \;= \; |\xi(8,1)| \frac{1}{2}^+|ggb>  \\
&=& \frac{1}{\sqrt{6}}[uds - usd + sud - sdu + dsu - dus] \otimes \\
&& \frac{1}{3\sqrt{2}} [ (2 g_{\uparrow} g_{\uparrow} b_{\downarrow} 
- g_{\uparrow} g_{\downarrow} b_{\uparrow} - g_{\downarrow} g_{\uparrow} b_{\uparrow} )
+ Perms(1 \leftrightarrow 3, 2 \leftrightarrow 3)] \nonumber \\
\nonumber \\
\Lambda_c^0 &=&|\Lambda_c^0(\frac{1}{2}^+)|0,0,0> \;= \; |\xi(8,1)| \frac{1}{2}^+|rbg(0)>  \\
&=& \frac{1}{\sqrt{6}}[uds - usd + sud - sdu + dsu - dus] \otimes \\
&& \frac{1}{2\sqrt{3}} [ (r_{\uparrow} b_{\uparrow} g_{\downarrow} 
+ r_{\uparrow} b_{\downarrow} g_{\uparrow} + r_{\downarrow} b_{\uparrow} g_{\uparrow} )
+ Perms(1 \leftrightarrow 2, 2 \leftrightarrow 3, 1 \leftarrow 3)] \nonumber \\
\end{eqnarray}
The color octet states are named according to their flavor octet counterparts. In this naming
convention, the $+,0,-$ superscripts refer to the {\it effective $U(1)$ color charge} of the state,
not the electric charge (which is zero for any flavor singlet).  To be explicit, we write the octet  
{\it color electric charge (CEC)} operator $\hat{Q}^c_i$ (for $i$ = quark index):
\begin{equation}
\hat{Q}^c_i \;=\; (\hat{I}^c_3 \;+\; \frac{1}{2}\hat{Y}^c)_i \;,
\end{equation}
which is the color generalization of the $U(1)$ electric charge quantization of $SU(3)$ flavor
octet states. The color isospin, hypercharge, and CEC quantum numbers are natural analogs of the
corresponding $SU(3)$ flavor quantum numbers. With the C/F transformation previously introduced,
we have the following quark quantum number assignments for color isospin, hypercharge, and CEC
as $|color(I^c,I^c_3,Y^c|Q^c)>$:
\begin{eqnarray}
&& |r\;(\frac{1}{2},\frac{1}{2},\frac{1}{3}|\frac{2}{3})> \\
&& |b\;(\frac{1}{2},-\frac{1}{2},\frac{1}{3}|-\frac{1}{3})> \\
&& |g\;(0,0,-\frac{2}{3}|-\frac{1}{3})> \;,
\end{eqnarray}
with the corresponding algebra:
\begin{eqnarray}
\hat{Q}^c \; |r> &=& \frac{2}{3} \; |r> \;, \\
\hat{Q}^c \; |b> &=& -\frac{1}{3} \; |b> \;, \\
\hat{Q}^c \; |g> &=& -\frac{1}{3} \; |g> \;, \; etc. 
\end{eqnarray}
{\bf Color confinement can be expressed as the condition that all physical hadrons must have total 
color electric charge equal to zero (and overall color antisymmetry).} For all hadrons (including
exotic multiquark systems) the total color electric charge is the sum over all quark contributions:
$Q^c = \sum_i Q^c_i$.

		\subsubsection{Color Octet $\alpha$(u\sbar) States}

Next consider the $\alpha(q\bar{s})|8_A,2_A|0^-,1^->$ antisymmetric color-octet and flavor-doublet
states $|\alpha^{Q^c}_c(spin^{parity})|I^c,I^c_3,Y^c>$: \\
\begin{eqnarray}
K_c^+ &=&|K_c^+(0^-)|\frac{1}{2},\frac{1}{2},1>  \\
&=& \frac{1}{\sqrt{2}}( |r \bar{g}> - |\bar{g} r>) \otimes 
\frac{1}{\sqrt{2}}( \uparrow \downarrow - \downarrow \uparrow) \otimes 
\frac{1}{\sqrt{2}} (q \bar{s} - \bar{s} q)  \; ,\\
\nonumber \\
K_c^0&=&|K_c^0(0^-)|\frac{1}{2},-\frac{1}{2},1> \\
&=& \frac{1}{\sqrt{2}}( |b \bar{g}> - |\bar{g} b>) \otimes 
\frac{1}{\sqrt{2}}( \uparrow \downarrow - \downarrow \uparrow) \otimes 
\frac{1}{\sqrt{2}} (q \bar{s} - \bar{s} q)  \; ,\\
\nonumber \\
\pi_c^+&=&|\pi_c^+(0^-)|1,1,0>  \\
&=& \frac{1}{\sqrt{2}}( |r \bar{b}> - |\bar{b} r>) \otimes 
\frac{1}{\sqrt{2}}( \uparrow \downarrow - \downarrow \uparrow) \otimes 
\frac{1}{\sqrt{2}}(q \bar{s} - \bar{s} q)  \; ,\\
\nonumber \\
\pi_c^0  &=& |\pi_c^0(0^-)|1,0,0> \\
&=& \frac{1}{\sqrt{2}} \left( \frac{1}{\sqrt{2}}(|r \bar{r}> - |\bar{r} r>) - 
\frac{1}{\sqrt{2}}(|b \bar{b}> - |\bar{b} b>) \right) \\ 
&\otimes& 
\frac{1}{\sqrt{2}}( \uparrow \downarrow - \downarrow \uparrow) 
 \otimes \frac{1}{\sqrt{2}} (q \bar{s} - \bar{s} q)  \; , \nonumber \\
\nonumber \\
\pi_c^- &=& |\pi_c^-(0^-)|1,-1,0> \\
&=& \frac{1}{\sqrt{2}}( |b \bar{r}> - |\bar{r} b>) \otimes 
\frac{1}{\sqrt{2}}( \uparrow \downarrow - \downarrow \uparrow) \otimes 
\frac{1}{\sqrt{2}}(q \bar{s} - \bar{s} q)  \; ,\\
\nonumber \\
\bar{K}_c^0 &=&|\bar{K}_c^0(0^-)|\frac{1}{2},\frac{1}{2},-1> \\
&=& \frac{1}{\sqrt{2}}( |g \bar{b}> - |\bar{b} g>) \otimes 
\frac{1}{\sqrt{2}}( \uparrow \downarrow - \downarrow \uparrow) \otimes 
\frac{1}{\sqrt{2}} (q \bar{s} - \bar{s} q)  \; ,\\
\nonumber \\
\bar{K}_c^-&=&|\bar{K}_c^-(0^-)|\frac{1}{2},-\frac{1}{2},-1> \\
&=& \frac{1}{\sqrt{2}}( |g \bar{r}> - |\bar{r} g>) \otimes 
\frac{1}{\sqrt{2}}( \uparrow \downarrow - \downarrow \uparrow) \otimes 
\frac{1}{\sqrt{2}} (q \bar{s} - \bar{s} q)  \; ,\\
\nonumber \\
\eta_c^0 &=& |\eta_c^8(0^-)|0,0,0> \\
&=& \frac{1}{2\sqrt{3}} \left( (|r \bar{r}> - |\bar{r} r>) + 
(|b \bar{b}> - |\bar{b} b>) ) - 2( |g \bar{g}> - |\bar{g} g> ) \right)
\\
&\otimes& \frac{1}{\sqrt{2}}( \uparrow \downarrow - \downarrow \uparrow) \otimes 
\frac{1}{\sqrt{2}} (q \bar{s} - \bar{s} q)  \; , \nonumber
\end{eqnarray}
where $q = u$ is the positive electric charge $I_3 = +\frac{1}{2}$, and $q = d$ is the negative electric
charge $I_3 = -\frac{1}{2}$ state. As with the color octet $\beta(B=1)$ states, the color octet members
of the $\alpha(B=0)$ cluster are named in correspondence with their flavor octet counterparts. The spin
one $\alpha$-states are spin triplets having the same color and flavor structure
$|\alpha^{Q^c}_c(spin^{parity})|I^c,I^c_3,Y^c>$: \\
\begin{eqnarray}
K_c^{*+} &=& |K_c^{*+}(1^-)|\frac{1}{2},\frac{1}{2},1> \\
&=& \frac{1}{\sqrt{2}}( |r \bar{g}> - |\bar{g} r>) \otimes 
\left[(\uparrow \uparrow),\frac{1}{\sqrt{2}}( \uparrow \downarrow + \downarrow \uparrow),
(\downarrow \downarrow)\right] \otimes 
\frac{1}{\sqrt{2}}(q \bar{s} - \bar{s} q)  \;,\\
\nonumber \\
K_c^{0*} &=& |K_c^{*0}(1^-)|\frac{1}{2},-\frac{1}{2},1> \\
&=& \frac{1}{\sqrt{2}}( |b \bar{g}> - |\bar{g} b>) \otimes 
\left[(\uparrow \uparrow),\frac{1}{\sqrt{2}}( \uparrow \downarrow + \downarrow \uparrow),
(\downarrow \downarrow)\right] \otimes 
\frac{1}{\sqrt{2}}(q \bar{s} - \bar{s} q), \;\;\;\;\; \\
&& {\bf \vdots} \nonumber
\end{eqnarray}
etc.~for the other octet members. 

		\subsubsection{Color Octet Pentaquark States}

With these $\alpha$(q\sbar) and $\beta$(uds) basis states, we now form the totally antisymmetric
$(\alpha - \beta)$ molecules. We proceed by coupling color, flavor, spin, and space wavefunctions
to overall quantum numbers of the physical $(\alpha-\beta)$ states. As previously discussed, overall
fermion antisymmetry in the $\beta$-cluster follows directly from the antisymmetry of the flavor
singlet $F_A^1(\beta)$. Construction of the symmetrized $F_S^2(\alpha \beta|X)$ product function
is straightforward yet tedious, so we omit the details. However, since antisymmetry of the pentaquark
color function $C_A^1(\alpha \beta|X)$ is crucial for color confinement, we show explicit details
of the construction. To proceed, we note that each member of the anti-color triplet
($\bar{r},\bar{b},\bar{g}$) is color-equivalent to the anti-symmetrized product of two color states:
\begin{eqnarray}
|\bar{r}> &=& \frac{1}{\sqrt{2}} \left( |b>|g> - |g>|b> \right) \\
|\bar{b}> &=& \frac{1}{\sqrt{2}} \left( |r>|g> - |g>|r> \right) \\
|\bar{g}> &=& \frac{1}{\sqrt{2}} \left( |b>|r> - |r>|b> \right) \;.
\end{eqnarray}

When the mixed symmetry color function from the $\beta$-cluster $C_{MS}^8(\beta)$ is coupled to
the mixed symmetry color function from the $\alpha$-cluster $C_{MS}^8(\alpha)$, the $SU(3)$
Clebsch-Gordon coefficients for many pairings will be zero because those products can not contribute
to a zero color charge state. For example, consider the coupling of a $(rrb)$ color configuration
in $\beta$ with a $(r \bar{g})$ state in $\alpha$. If we symbolically express the color
Clebsch-Gordon coefficient as 
$C_G(8_{\alpha} \otimes 8_{\beta} \rightarrow 1_{X}) = (\alpha||C^{\alpha}_1,\bar{C}^{\alpha}_2
||X^1|| C^{\beta}_1,C^{\beta}_2,C^{\beta}_3||\beta)$
then 
\begin{equation}
C_G\left(\alpha^8(r\bar{g}) \otimes \beta^8(rrb)|X^1 \right) \;=\; 
\left( \alpha||r,\bar{g}||X^1||r,r,b|| \beta \right) \;=\; 0 .
\end{equation}
This coupling must be zero (and thus $|rrb>\otimes |r\bar{g}>$ is prohibited) since the net color
electric charge of such a pentaquark state is not zero:
\begin{eqnarray}
\hat{Q^c}\; |\beta(rrb) \otimes \alpha(r \bar{g})> &=& 
\left( Q_r + Q_r + Q_b + Q_r + Q_{\bar{g}} \right) |\beta(rrb)\otimes \alpha(r \bar{g})> \\
&=& \left( \frac{2}{3} + \frac{2}{3} -\frac{1}{3} + \frac{2}{3} + \frac{1}{3} \right) |\beta(rrb)\otimes 
	\alpha(r \bar{g})> \;\;\;\;\; \\
&=& (2)\; |\beta(rrb)\otimes \alpha(r \bar{g})> \; \neq 0 .
\end{eqnarray}

Next consider a typical coupling between $\alpha - \beta$ color states that will contribute
to a color neutral $X$-state: 
\begin{equation}
C_G\left(\alpha^8(r\bar{g}) \otimes \beta^8(ggb)|X^1 \right) \;=\; 
\left( \alpha||r,\bar{g}||X^1||g,g,b||\beta \right) \; = \; \pm {\cal N}_X^c ,
\end{equation}
where ${\cal N}_X^c$ is the $X$-state pentaquark color normalization, and the sign will depend on
the adopted phase convention. Because the pentaquark color singlet wavefunction must have overall
$A$-symmetry, the normalization factor must have the same magnitude for all terms in the antisymmetric
superposition. There are 9-primary (non-symmetrized) zero color combinations (only 8 independent)
with non-zero $8 \otimes 8 \rightarrow 1$ Clebsch-Gordon coefficients: 
$(g \bar{r}|rrb)$, $(r \bar{g}|ggb)$, $(b \bar{g}|ggr)$, $(b \bar{r}|rrg)$, $(g \bar{b}|bbr)$,
$(r \bar{b}|bbg)$, $(r \bar{r}|rbg)$, $(b \bar{b}|rbg)$, and $(g \bar{g}|rbg)$. 
Explicit color antisymmetry can be constructed in a sequence of color transformations starting
with an arbitrary primary pentaquark color combination. For example, starting with the 
$(\alpha_{i=1}|\beta_{i=1})=(g \bar{r}|rrb)$ term, we first write the product of (non-normalized) 
MS-$|\alpha>$ and MS-$|\beta>$ color octet functions:
\begin{equation}
C_{MS}^8(\alpha_1|g \bar{r}|X) \; C_{MS}^8(\beta_1|rrb|X)
\;=\;(g \bar{r} - \bar{r}g)\;(rrb - rbr + brr) \;,
\end{equation}
and likewise for the 8-independent products. Next transform this first term by applying the color
transformation $r \leftrightarrow b$ (and subtract to anti-symmetrize):
\begin{equation}
 C_{MS}^8(g \bar{b}|X)\; C_{MS}^8(bbr|X)\;=\; (g \bar{r} - \bar{r}g)\;(bbr - brb + rbb)\;.
\end{equation}
Repeated application of the various color transformations: $r \leftrightarrow b$,
$r \leftrightarrow g$, $b \leftrightarrow g$ generates the following (normalized) antisymmetric 
{\it color cluster product expansion}:
\begin{equation}
C_A^1(\alpha \beta|X) \;=\;
{\cal N}_X^c \; \sum_{i=1}^{8} (-1)^{i+1}\; C_{MS}^8(\alpha_i|X) \; C_{MS}^8(\beta_i|X)  \;, 
\end{equation}
where the $|\alpha>$-color functions are:
\begin{eqnarray}
&& C_{MS}^8(\alpha_1|g\bar{r}|X) = (g\bar{r}-\bar{r}g) \\
&& C_{MS}^8(\alpha_2|g\bar{b}|X) = (g\bar{b}-\bar{b}g) \\
&& C_{MS}^8(\alpha_3|r\bar{b}|X) = (r\bar{b}-\bar{b}r) \\
&& C_{MS}^8(\alpha_4|b\bar{r}|X) = (b\bar{r}-\bar{r}b) \\
&& C_{MS}^8(\alpha_5|r\bar{g}|X) = (r\bar{g}-\bar{g}r) \\
&& C_{MS}^8(\alpha_6|b\bar{g}|X) = (b\bar{g}-\bar{g}b) \\
&& C_{A}^8(\alpha_7|r \bar{r},b\bar{b},g\bar{g}|X) = 
(r \bar{r} - \bar{r}r + b\bar{b} - \bar{b}b + g\bar{g} - \bar{g}g) \\
&& C_{S}^8(\alpha_8|r\bar{r},b\bar{b},g\bar{g}|X) = 
(r \bar{r} + \bar{r}r + b\bar{b} + \bar{b}b + g\bar{g} + \bar{g}g) 
\end{eqnarray}
and the $|\beta>$-color functions are: 
\begin{eqnarray}
&& C_{MS}^8(\beta_1|rrb|X) = (rrb -rbr+brr) \\
&& C_{MS}^8(\beta_2|bbr|X) = (bbr -brb+rbb) \\
&& C_{MS}^8(\beta_3|bbg|X) = (bbg -bgb+gbb) \\
&& C_{MS}^8(\beta_4|rrg|X) = (rrg -rgr+grr) \\
&& C_{MS}^8(\beta_5|ggb|X) = (ggb -gbg+bgg) \\
&& C_{MS}^8(\beta_6|ggr|X) = (ggr -grg+rgg) \\
&& C_{S}^8(\beta_7|rbg|X) = (rbg+rgb+grb+gbr+bgr+brg) \\
&& C_{A}^8(\beta_8|rbg|X) = (rbg-rgb+grb-gbr+bgr-brg)  
\end{eqnarray}
Note within both the $\alpha$ and $\beta$ octets there are two terms that have zero net color 
charge. These color neutral octet member could be written with $MS$-symmetry, but to make the
Clebsch-Gordon factors simple (and to demonstrate explicit $A$-color symmetry) we show the two
independent $S$ and $A$-symmetry combinations. After the sum in Eq.(66) over the 8-octet states
is performed, the color function will be totally antisymmetric in color interchange. This means that
the interchange of any two colors will generate the opposite overall sign in the color cluster product
expansion. For example, under the color exchange $r \leftrightarrow b$:
\begin{eqnarray}
&& C_{MS}^8(\alpha_1|g\bar{r} |X) \; C_{MS}^8(\beta_1|rrb|X) \; \leftrightarrow (-)
C_{MS}^8(\alpha_2|g\bar{b}|X) \; C_{MS}^8(\beta_2|bbr|X)\; \\
&& C_{MS}^8(\alpha_3|r\bar{b} |X)\; C_{MS}^8(\beta_3|bbg|X) \; \leftrightarrow (-)
C_{MS}^8(\alpha_4|b\bar{r}|X) \; C_{MS}^8(\beta_4|rrg|X)\; \\
&& C_{MS}^8(\alpha_5|r\bar{g} |X)\; C_{MS}^8(\beta_5|ggb|X) \; \leftrightarrow (-)
C_{MS}^8(\alpha_6|b\bar{g}|X) \; C_{MA}^8(\beta_6|ggr|X)\; \\
&& C_{A}^8(\alpha_7| r \bar{r},b\bar{b},g\bar{g}|X) \; C_{S}^8(\beta_7|rbg|X) 
\; \leftrightarrow (-)
C_{A}^8(\alpha_7| b \bar{b},r\bar{r},g\bar{g}|X)\; C_{S}^8(\beta_7|brg|X) \; \\
&& C_{S}^8(\alpha_8|r \bar{r},b\bar{b},g\bar{g}|X) \; C_{A}^8(\beta_8|rbg|X)
\;\leftrightarrow (-)
C_{S}^8(\alpha_8|b \bar{b},r\bar{r},g\bar{g}|X) \; C_{A}^8(\beta_8|brg|X)\;\;\;\;\;\;\;\;\;
\end{eqnarray}
The QMM pentaquark color antisymmetry is now explicit. There are a total of 
$(6\times 2 \times 3) + (2 \times 6 \times 6) = 108$ terms in the antisymmetric sum, hence the
color function normalization is: ${\cal N}_X^c = \frac{1}{\sqrt{108}} = \frac{1}{6 \sqrt{3}}$.

	\subsection{Color Magnetic Moments and Confinement} 

We now turn our attention to a dynamical constraint that has two novel consequences: (1) to provide a
selection rule that limits the QMM excitation spectrum (allowing only 4-physical states); (2) to
provide dynamical relations between constants of the QMM Hamiltonian such that our calculated
pentaquark baryon mass spectrum has only one free parameter (the $\xi$ mass). Using the {\it color
electric charge (CEC)} operator described in the previous subsection, we now define the {\it color
magnetic moment (CMM)} operator:
\begin{equation}
\hat{\mu}^c \;=\; \sum_i \; \hat{\mu}^c_i \; \hat{\sigma}_{3 i} \;=\; \mu_0^c 
\sum_i \; \hat{Q}^c_i \; \hat{\sigma}_{3 i} \;,
\end{equation}
which is the color generalization of the $U(1)$ magnetic moment operator of the conventional
(non-exotic) quark model. The sum extends over all quarks and the color magnetic moment operator
$\hat{\mu}^c_i$ satisfies the $SU(3)$ algebra:
\begin{eqnarray}
\hat{\mu}^c |r> &=& \mu_0^c \; \hat{Q}^c \;|r> \;=\; \;\;\;
\frac{2}{3} \; \mu_0^c \; |r> \;=\; \mu_r^c \; |r>  \; , \\
\hat{\mu}^c |b> &=& \mu_0^c \; \hat{Q}^c \;|b> \;=\; 
-\frac{1}{3} \; \mu_0^c \; |b> \;=\; \mu_b^c \; |b>  \;, \\
\hat{\mu}^c |g> &=& \mu_0^c \; \hat{Q}^c \;|g> \;=\;   
-\frac{1}{3} \; \mu_0^c \; |g> \;=\; \mu_g^c \; |g>  \;.
\end{eqnarray}
The constant $\mu_0^c$ is the fundamental color magnetic moment, which may be regarded as a free
parameter of the QMM, however we take a different approach since its numerical value is not used
in our spectrum calculations. The important point about the constant $\mu_0^c$ is that its value
(proportional to the strong coupling constant $g_s = \sqrt{4 \pi \alpha_s}$) must be the same for
all quark colors to satisfy exact $SU(3)$ color symmetry.
 
We can now state our hypothesis of {\it color magnetic confinement (CMC)}:
\begin{center}
\fbox{
	{\bf The net color magnetic moment is zero for all physical hadrons.} 
}
\end{center}

If the CMC hypothesis is true, then the converse must also be true:
\begin{center}
\fbox{
	{\bf Any hadron with non-zero color magnetic moment is unphysical.} 
}
\end{center}

To see why the CMC conjecture is reasonable, we apply our hypothesis to a conventional quark 
model flavor octet state (the proton). First we take the direct product of standard $A$-color and
$MS$-spin functions for the proton:
\begin{eqnarray}
|p: \uparrow>_{_{2-3}} &=& \frac{1}{\sqrt{6}}\left( rbg-rgb+bgr-brg+grb-gbr \right) \\
&\times& \frac{1}{\sqrt{2}} \uparrow \left(\uparrow \downarrow - \downarrow \uparrow \right)\;,
\end{eqnarray}
where the (2-3) subscript denotes $A$-symmetry for interchange between the second and third quarks 
in the spin function (and no other pairwise interchange symmetry).
Obviously, the proton has CEC equal to zero. After applying the CMM operator, there are a total of
36 terms (12 terms with 3 quarks each) with exact cancellation between terms:
\begin{equation}
<\uparrow : p|\; \hat{\mu}^c \; |p: \uparrow> \;=\;  
<\uparrow : p|\; \mu_0^c \; \sum_i \; \hat{Q}^c_i \; \hat{\sigma}_{3 i} \; |p: \uparrow> 
 \;=\; 0 \;.
\end{equation}
Note the cancellation is only exact if $\mu_0^c$ is the same for each color. The physical proton is
built from a superposition of $MS$-symmetry flavor functions, but the exact cancellation in the
CMM holds separately for each one, so the proton satisfies the CMC hypothesis. We can show that
every member of the (non-exotic) baryon ($\frac{1}{2}^+$) octet, ($\frac{3}{2}^+$) decuplet and
meson ($0^-,1^-$) nonets also satisfy the CMC hypothesis (but omit the details here). It is possible
that some (or many) of the excited states of the conventional (non-exotic) quark model may not
satisfy the generalized CMC sum rule (including orbital CMM contributions) and therefore drop out of
the excitation spectrum. The CMC principle may help to resolve the longstanding missing nucleon ($N^*$) 
resonance problem of baryon spectroscopy, or CMC may simply not be generally true, however we leave
this question open for future consideration. We now proceed to apply the CMC conjecture to constrain
the Hamiltonian of our pentaquark baryon molecule states. 

	\subsection{Color Magnetic Confinement (CMC) Constraint}

The complication of applying CMC in the QMM pentaquark basis is that there must be at least one unit
of relative orbital angular momentum between color octet $|\alpha(q\bar{s})>$ and $|\beta(uds)>$
clusters. Hence, there should be a CMM contribution from the colorful orbital motion of the molecular
state. If CMC is generally true, then we must calculate the orbital contribution to the total molecular
color magnetic moment. When this is accomplished, there will be consistency relations between the
color spin-spin and spin-orbit coupling strengths in our QMM Hamiltonian so that overall CMC is satisfied. 

Our molecular pentaquark spectrum is generated by the following Hamiltonian:
\begin{eqnarray}
&& \hat{H} \;=\; \hat{H}_0 + 
C_F^8 \left(\frac{\alpha_s}{r}\right) + \frac{A_{\alpha \beta}}{M_\alpha M_\beta} 
\vec{S}_{\alpha} \cdot \vec{S}_{\beta} \left(\frac{1}{r} \frac{dV}{dr} \right) 
+ \frac{B_{\alpha \beta}}{M_\alpha M_\beta} \vec{L} \cdot \vec{S} 
\left(\frac{1}{r}\frac{dV}{dr} \right)  \\
&& \hat{H}_0 \;=\; - \frac{\nabla^2}{2 M_{\alpha \beta}} 
+ \frac{1}{2}  k  r^2 \; \\
&& V(r) \;=\; C_F^8 \left(\frac{\alpha_s}{r}\right) + \frac{1}{2}  k  r^2 \; \\
&& \left(\frac{1}{r} \frac{dV}{dr} \right) \;=\; 
-C_F^8 \left(\frac{\alpha_s}{r^3}\right) +   k   \;
\end{eqnarray}
where $\vec{S}= \vec{S}_\alpha + \vec{S}_\beta$ is the total spin, and $\vec{L}$ is the orbital angular
momentum operator. This Hamiltonian has the same form as a non-relativistic reduction of the Dirac equation
with a minimally coupled gauge field 4-vector potential (generating electric and magnetic interactions).
Cluster confinement is enforced by the harmonic oscillator potential, which generates a convenient S.H.O.
basis of radial wavefunctions for our spectrum calculations. As usual, $\alpha_s$ is the strong coupling
constant of QCD~\cite{Griffiths87} (fixed by conventional quark model spectroscopy $\alpha_s = 0.3$). We
checked that using the running coupling $\alpha_s(r)$ gives an effect less than 1\% in the radial
expectation values calculated in this work. The factor $C_F^8$ multiplying the Coulomb term is the color
factor of one gluon exchange between color octet quark clusters~\cite{Griffiths87} (derived later). The
$A_{\alpha \beta}$ and $B_{\alpha \beta}$ coefficients are combinations of {\it color gyromagnetic ratio}
$\gamma$-factors (analogous to the definition in electrodynamics) corresponding to the color magnetic
moments of the quark clusters and orbital motion, which are constrained and related by the CMC sum rule.

We proceed by defining the total color magnetic moment of the molecular pentaquark system $\vec{\mu}_{tot}$
in terms of the intrinsic $\vec{\mu}_{\alpha}$ and $\vec{\mu}_{\beta}$ cluster contributions, and the
orbital contribution $\vec{\mu}^{c}_L$:
\begin{eqnarray}
\vec{\mu}^c_{tot} &=&  \vec{\mu}^c_\alpha \;+\; \vec{\mu}^c_\beta \;+\;
\vec{\mu}^{c}_L \;.
\end{eqnarray}
The generalized CMC constraint is: 
\begin{center}
\fbox{\bf CMC: $\vec{\mu}^c_{tot} = \vec{0}$ } \\
\end{center}

\begin{figure}[!hbt]
\begin{center}
\mbox{
\epsfxsize = 8cm
\epsfysize = 8cm
\epsffile{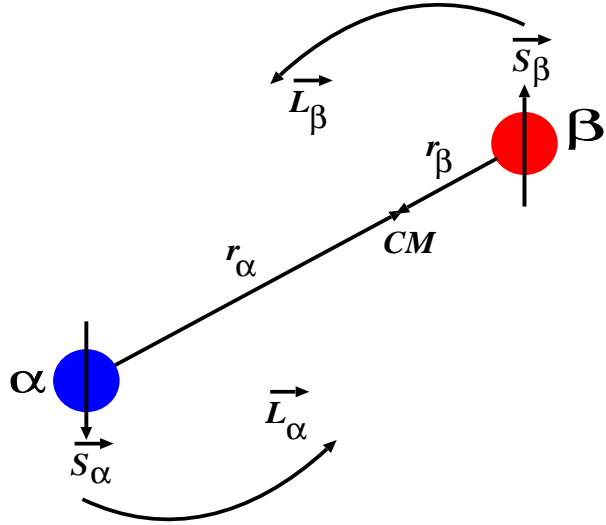}
}
\end{center}
	\caption{The center-of-mass orbital angular momentum of the pentaquark system.}
	\label{fig:LSorbit}
\end{figure}

The intrinsic $\alpha$ and $\beta$ magnetic moments are expressed:
\begin{eqnarray}
\vec{\mu}^c_\alpha &=& \gamma_\alpha \; \left(\frac{g_s}{2M_\alpha}\right) \; 
 \vec{S}_\alpha \; \\
\vec{\mu}^c_\beta &=&  \gamma_\beta \; \left(\frac{g_s}{M_\beta}\right) \; 
 \vec{S}_\beta \;
\end{eqnarray}
where $g_s = \sqrt{4 \pi \alpha_s}$ is the QCD strong coupling constant, and the
($\gamma_\alpha$,$\gamma_\beta$) factors are dimensionless color gyromagnetic ratio factors of the
($\alpha$,$\beta$) clusters. It is possible (yet tedious) to relate the color gyromagnetic ratio constants
to the fundamental color magnetic moment $\mu_0^c$ of the CMM quark operator using the explicit cluster
basis wavefunctions of section~\ref{sec:wavefunctions} (analogous to how traditional quark model $SU(3)_F$
baryon magnetic moments are related to quark magnetic moments). However, we apply the generalized CMC rule
to constrain the effective molecular CMM constants, and hence do not pursue these QMM parameter relations
in this work. Note the baryon $\beta$ cluster does not have a factor of 2 in the mass denominator because
we assume the Landau $g=2$ factor of Dirac electrodynamics which enhances fermion magnetic fields relative
to non-fermion fields (such as those generated by $|\alpha>$). Next, we define the intrinsic
($\alpha \beta$)-molecule CMM $\vec{\mu}^c_{\alpha \beta}$ and its corresponding effective color
gyromagnetic ratio $\gamma^{\alpha \beta}_S$ by coupling to total spin:
$\vec{S} = \vec{S}_\alpha + \vec{S}_\beta$:
\begin{eqnarray}
\vec{\mu}^c_{\alpha \beta} &=& \vec{\mu}^c_\alpha \;+\; \vec{\mu}^c_\beta \;=\; 
(-)^{J-S} \; \gamma^{\alpha \beta}_S \; 
\left(\frac{g_s}{M_{\alpha \beta}}\right) \; \vec{S}\;,
\end{eqnarray}
where $M_{\alpha \beta} = \frac{M_\alpha M_\beta}{M_\alpha + M_\beta}$ is the ($\alpha$-$\beta$) reduced
mass, and the $J$-dependent phase is required for a technical reason we come to shortly. The orbital color
magnetic moment $\vec{\mu}_L^c$ is expressed in terms of the total orbital angular momentum $\vec{L}$ and 
effective orbital gyromagnetic ratio $\gamma_L^{\alpha \beta}$:
\begin{eqnarray}
\vec{\mu}^c_L &=&
\gamma_L^{\alpha \beta}
\; \left(\frac{g_s}{2 M_{\alpha \beta}}\right) \vec{L} \;. 
\end{eqnarray}
The origin of the orbital gyromagnetic ratio factor can be traced to the fact that the total orbital color
magnetic moment is generated asymmetrically between the colorful clusters of different mass. Each cluster
carries a different amount of orbital angular momentum relative to the center of mass of the molecular
system (Fig.~\ref{fig:LSorbit}). 

For orbiting ($\alpha$-$\beta$) clusters in a particular spin configuration, each carries individual
angular momentum:
\begin{eqnarray}
\vec{L}_\alpha &=& \vec{r}_\alpha \times \vec{p}_\alpha 
\;{classical \over \longrightarrow}\; 
\left(M_\alpha \;r^2_\alpha \;\omega_l \right) \; \hat{z} \\
\vec{L}_\beta &=& \vec{r}_\beta \times \vec{p}_\beta 
\;{classical \over \longrightarrow}\; 
\left(M_\beta \;r^2_\beta \;\omega_l \right) \; \hat{z} 
\end{eqnarray} 
where $\hat{z}$ is the quantization axis of spin and orbital angular momentum, and the angular
velocity (expectation value) $\omega_l$ must be the same for both clusters so the center of mass
position is fixed in space (i.e., to satisfy conservation of linear momentum with no external forces).
Also, the quantum two-body system satisfies Newton's $3^{rd}$ Law (with expectation values used for
radial factors): 
\begin{equation} 
M_\alpha \; r_\alpha \; \omega_l^2 \;=\; M_\beta \; r_\beta \; \omega_l^2 \;.
\end{equation}
Total orbital angular momentum is the vector sum of both cluster contributions, and naturally
generates the standard reduced mass $M_{\alpha \beta}$  factor of a CM separated two body system: 
\begin{eqnarray}
\vec{L} &=& \vec{L}_\alpha \;+\; \vec{L}_\beta \\
&=& \left( M_\alpha \;r^2_\alpha \;\omega_l \;+\; M_\beta \;r^2_\beta \;\omega_l \right) \; 
\hat{z} \\
&=& M_{\alpha \beta} \; (r_\alpha + r_\beta)^2 \; \omega_l \; \hat{z} \;.
\end{eqnarray}
Because of Newton's $3^{rd}$ Law, Eq.~(106), each cluster angular momentum is proportional to the
total angular momentum, depending on mass ratio factors:
\begin{eqnarray}
\vec{L}_\alpha &=& \left(\frac{M_\beta}{M_\alpha}\right) \; \vec{L}_\beta \\
\vec{L}_\alpha &=&  \frac{1}{[1+(M_\alpha/M_\beta)]} \;
\vec{L} \\
\vec{L}_\beta &=& \left(\frac{M_\alpha}{M_\beta}\right) 
\frac{1}{[1+(M_\alpha/M_\beta)]} \; \vec{L} \;.
\end{eqnarray}
Returning to the orbital color gyromagnetic ratio, we couple the separate $\alpha$ and $\beta$ orbital
CMM contributions to the total orbital angular momentum:
\begin{equation}
\gamma_L^{\alpha \beta} \left(\frac{g_s}{2 M_{\alpha \beta}}\right) \vec{L} \;=\; 
\left(\frac{g_s}{2 M_{\alpha}}\right) \vec{L}_{\alpha} \;+\; 
\left(\frac{g_s}{2 M_{\beta}}\right) \vec{L}_{\beta} \;.
\end{equation}
Using Eqs.~(111,112) we obtain:
\begin{equation}
\gamma_L^{\alpha \beta} \;=\; \frac{M_\alpha M_\beta}{(M_\alpha + M_\beta)^2} 
\left[\frac{M_\alpha}{M_\beta} \;+\; \frac{M_\beta}{M_\alpha}\right] \;.
\end{equation}
It is interesting to note the equal mass and infinite asymmetric mass limits:
\begin{eqnarray}
\lim_{M_\alpha = M_\beta} \; \gamma_L^{\alpha \beta} &\rightarrow & \frac{1}{2} \\
\lim_{M_\alpha >> M_\beta} \; \gamma_L^{\alpha \beta} &\rightarrow & 1 \\
\lim_{M_\beta >> M_\alpha} \; \gamma_L^{\alpha \beta} &\rightarrow & 1 \;.
\end{eqnarray}

We now explore the effective color gyromagnetic ratio of the coupled molecular spin CMM,
$\gamma_{S}^{\alpha \beta}$. Using Eqs.~(100,101,102), we have:
\begin{equation}
(-)^{J-S} \; \gamma_{S}^{\alpha \beta} \; \left(\frac{g_s}{M_{\alpha \beta}}\right) \; \vec{S} \;=\;
\gamma_{\alpha} \left( \frac{g_s}{2 M_\alpha}\right) \vec{S}_\alpha \;+\; 
\gamma_{\beta} \left( \frac{g_s}{M_\beta}\right) \vec{S}_\beta  \;.
\end{equation}
The $J$-dependent phase factor is required for consistency because in the coupled 
quantum mechanical molecular system, $(S_\alpha,S_\beta,S,S_z,L,L_z,J,J_z)$ are ``good'' quantum numbers
(operators commute with Hamiltonian), but $(S_{\alpha z},S_{\beta z})$ are not. To obtain the required
relations, first examine the classical $+\hat{z}$ projection of Eq.~(118)
with $(S = \frac{1}{2}|S_{\alpha}=0,S_\beta = \frac{1}{2})$:
\begin{equation} 
(-)^{J-\frac{1}{2}} \; \gamma_{S}^{\alpha \beta}(\frac{1}{2}|0,\frac{1}{2}) \; 
\left(\frac{g_s}{M_{\alpha \beta}}\right) \; 
(+\frac{1}{2}) \;=\;
\gamma_\alpha \; \left(\frac{g_s}{2 M_\alpha}\right) (0) \;+\; 
\gamma_{\beta} \left( \frac{g_s}{M_\beta}\right) (+\frac{1}{2}) \;.  
\end{equation}
Our first consistency relation for $\gamma_S^{\alpha \beta}(S|S_\alpha,S_\beta)$ follows directly:
\begin{equation}
(-)^{J-\frac{1}{2}} \; \gamma_{S}^{\alpha \beta}(\frac{1}{2}|0,\frac{1}{2}) \;=\;
\left(\frac{M_{K_c \xi_c}}{M_{\xi_c}}\right) \; \gamma_{\xi_c} \;,
\end{equation}
where $\alpha = K_c(q \bar{s}|0^-)$ (a spin $0^-$ color octet cluster with kaon spin-flavor composition),
and $\beta = \xi_c(uds|\frac{1}{2}^+)$, the previously discussed color octet, flavor singlet baryon
cluster state. Because the $K_c(0^-)$ cluster has spin zero, its color gyromagnetic ratio must equal
zero: $\gamma_{K_c} = 0$.

Next consider a spin one (color octet) meson cluster ($S_\alpha = 1$), the
$\alpha = K^*_c(q \bar{s}|1^-)$ with vector kaon spin-flavor composition coupled with the
$\beta = \xi_c(uds|\frac{1}{2}^+)$:  
\begin{equation}
(-)^{J-\frac{1}{2}} \; \gamma_{S}^{\alpha \beta}(\frac{1}{2}|1,\frac{1}{2}) \;=\;
M_{K^*_c \xi_c} \left( \frac{\gamma_{K^*_c}}{M_{K^*_c}} \;-\; 
\frac{\gamma_{\xi_c}}{M_{\xi_c}}\right) \;.
\end{equation}
The last possible molecular spin combination is the vector $\alpha = K^*_c(q \bar{s}|1^-)$ coupled to
the $\beta=\xi_c(uds|\frac{1}{2}^+)$ in a $S = \frac{3}{2}$ configuration:
\begin{equation}
(-)^{J-\frac{3}{2}} \; \gamma_{S}^{\alpha \beta}(\frac{3}{2}|1,\frac{1}{2}) \;=\;
(\frac{1}{3}) \; M_{K^*_c \xi_c} \left( \frac{\gamma_{K^*_c}}{M_{K^*_c}} \;+\; 
\frac{\gamma_{\xi_c}}{M_{\xi_c}}\right) \;.
\end{equation}

Finally, we apply the CMC constraint to develop consistency relations between the effective gyromagnetic
ratios of the cluster states ($\gamma_{K_c}$,$\gamma_{K^*_c}$,$\gamma_\xi$) and the orbital factors.
Eq.~(99) can be written as:
\begin{eqnarray}
\vec{\mu}^c_{tot} \;=\; \vec{\mu}_{\alpha \beta}^c \;+\; \vec{\mu}_L^c &=& 
(-)^{J-S} \; \gamma_{S}^{\alpha \beta} \; \left(\frac{g_s}{M_{\alpha \beta}}\right) \; \vec{S} 
\;+\; \gamma_{L}^{\alpha \beta} \; \left(\frac{g_s}{2 M_{\alpha \beta}}\right) \; \vec{L} \\
&=&
\gamma_{(J|LS)}^{\alpha \beta} \; \left( \frac{g_s}{M_{\alpha \beta}} \right) \; \vec{J} \;
\;\;\; {CMC \over \longrightarrow} \;\;\;\; 0 \;.
\end{eqnarray}
So the CMC condition requires the color gyromagnetic ratio of any physical molecular state with quantum
numbers $(J,J_z|L,L_z,S,S_z)$ to be identically zero: $\gamma_{(J|LS)}^{\alpha \beta} = 0$. Examining the
$\hat{z}$ projections for the different $(J|LS)$ quantum couplings, we find the following consistency
relations for $\alpha \beta(J|LS)$:
\begin{eqnarray}
&& K_c \xi (\frac{1}{2}|1,\frac{1}{2}): \;\;\; \gamma_{S}^{\alpha \beta}(\frac{1}{2}|0,\frac{1}{2})
	= \gamma_L^{K_c \xi} \\
&& K_c \xi (\frac{3}{2}|1,\frac{1}{2}): \;\;\;  -\gamma_{S}^{\alpha \beta}(\frac{1}{2}|0,\frac{1}{2})
	= -\gamma_L^{K_c \xi} \\
&& K^*_c \xi (\frac{1}{2}|1,\frac{1}{2}): \;\;\; \gamma_{S}^{\alpha \beta}(\frac{1}{2}|1,\frac{1}{2})
	= \gamma_L^{K^*_c \xi} \\
&& K^*_c \xi (\frac{3}{2}|1,\frac{1}{2}): \;\;\; -\gamma_{S}^{\alpha \beta}(\frac{1}{2}|1,\frac{1}{2})
	= -\gamma_L^{K^*_c \xi} \\
&& K^*_c \xi (\frac{1}{2}|1,\frac{3}{2}): \;\;\; -\gamma_{S}^{\alpha \beta}(\frac{3}{2}|1,\frac{3}{2})
	= 
(\frac{1}{3}) \gamma_L^{K^*_c \xi} \\
&& K^*_c \xi (\frac{3}{2}|1,\frac{3}{2}): \;\;\; \gamma_{S}^{\alpha \beta}(\frac{3}{2}|1,\frac{3}{2})
	= 0 \\
&& K^*_c \xi (\frac{5}{2}|1,\frac{3}{2}): \;\;\; -\gamma_{S}^{\alpha \beta}(\frac{3}{2}|1,\frac{3}{2})
	= -(\frac{1}{3}) \gamma_L^{K^*_c \xi} 
\end{eqnarray} 
Note the only consistent solution for Eqs.~(129,130,131) is: 
$\gamma_{S}^{\alpha \beta}(\frac{3}{2}|1,\frac{3}{2}) = 0$. But that would also require 
$\gamma_L^{K^*_c \xi} \rightarrow 0$. Since $\gamma_L^{K^*_c \xi}$ is uniquely determined by Eq.~(114)
(non-zero), the $K^*_c \xi(J|LS) = (\frac{1}{2}|1,\frac{3}{2}), (\frac{5}{2}|1,\frac{3}{2})$
states cannot consistently satisfy the CMC condition and are therefore unbound (unphysical states) in the QMM.
Careful inspection will show the sign difference between the $J=\frac{1}{2}$ and $J=\frac{3}{2}$ configurations
of the $L=1$ and $S=\frac{1}{2}$ $K_c \xi$ and $K^*_c \xi$ solutions correspond to the $J$-dependent phase
factors in Eqs.~(120,121). Now using Eqs.~(120,121,122), together with the two independent (CMC consistent)
solutions Eqs.~(125,127) (or Eqs.~(126,128)) we find the corresponding set of relations between the QMM
gyromagnetic ratio factors:
\begin{eqnarray}
\gamma_{S}^{\alpha \beta}(\frac{1}{2}|0,\frac{1}{2}) &=& \gamma_L^{K_c \xi_c} 
\;=\;  
\left(\frac{M_{K_c \xi_c}}{M_{\xi_c}}\right) \; \gamma_{\xi_c} \; \\
\gamma_{S}^{\alpha \beta}(\frac{1}{2}|1,\frac{1}{2}) &=& \gamma_L^{K^*_c \xi_c} 
\;=\; 
M_{K^*_c \xi_c} \left( \frac{\gamma_{K^*_c}}{M_{K^*_c}} \;-\; 
\frac{\gamma_{\xi_c}}{M_{\xi_c}}\right)\;,
\end{eqnarray}
which we invert to express the unknown (phenomenological) cluster color gyromagnetic ratio factors
$\gamma_{K^*_c}$ and $\gamma_{\xi_c}$ in terms of the derived orbital factors $\gamma_L^{\alpha \beta}$, and
colorful cluster masses $M_\alpha = (M_{K_c}, M_{K^*_c})$, $M_\beta = M_{\xi_c}$:
\begin{eqnarray} 
\gamma_{\xi_c} &=& 
\left(\frac{M_{\xi_c}+M_{K_c}}{M_{K_c}}\right) \; \gamma_L^{K_c \xi_c}  \\
\gamma_{K^*_c} &=& 
\left( \frac{M_{K^*_c}+M_{\xi_c}}{M_{\xi_c}}\right) \gamma_L^{K^*_c \xi_c} \;+\;
\frac{M_{K^*_c}}{M_{K_c}} \left(\frac{M_{K_c}+M_{\xi_c}}{M_{\xi_c}}\right)\gamma_L^{K_c \xi_c}  
\end{eqnarray}
It is important to stress here that these parameter relations were derived analyzing the classical
$\hat{z}$ projection of the coupled spin $S_z = S_{\alpha z} + S_{\beta z}$, but we have already noted that
$S_{\alpha z}$ and $S_{\beta z}$ are not good quantum numbers of the molecular system. We will refer to this
parameter set as the ``{\it classical approximation}'' for the QMM-CMC constrained gyromagnetic ratios.

However, we can derive a set of parameters consistent with the rules of quantum mechanics simply by taking
expectation values of the spin operators associated with Eq.~(118). Using the standard spin (and later orbital)
expectation values:
\begin{eqnarray}
&& <\vec{S}\cdot \vec{S}> \;\;\;\;\;= \; S(S+1) \\
&& <\vec{S_\alpha} \cdot \vec{S_\alpha}> \;=\; S_\alpha(S_\alpha+1) \\
&& <\vec{S_\beta} \cdot \vec{S_\beta}> \;=\; S_\beta(S_\beta+1) \\
&& <\vec{S_\alpha}\cdot \vec{S_\beta}> \;=\; 
\frac{1}{2}[S(S+1) - S_\alpha(S_\alpha+1) -  S_\beta(S_\beta+1) ] \\
&& <\vec{L}\cdot \vec{L}> \;=\; L(L+1) \\
&& <\vec{J}\cdot \vec{J}> \;=\; J(J+1) \\
&& <\vec{L}\cdot \vec{S}> \;=\; 
\frac{1}{2}[ J(J+1) - L(L+1) -  S(S+1) ] \;,
\end{eqnarray}
by taking the $\vec{S}\cdot \vec{S}$ projection of Eq.~(118), we obtain the general quantum consistent
parameter relations: 
\begin{eqnarray}
&& (-)^{J-S} \; \gamma_S^{\alpha \beta}(S|S_\alpha,S_\beta) \;=\; \\
&& \frac{M_{\alpha \beta}}{S(S+1)}\left[ S_\alpha(S_\alpha+1) (\frac{\gamma_\alpha}{2 M_\alpha})
+ S_\beta(S_\beta+1) (\frac{\gamma_\beta}{M_\beta}) 
+ <\vec{S}_\alpha \cdot \vec{S}_\beta> 
(\frac{\gamma_\alpha}{2 M_\alpha} + \frac{\gamma_\beta}{M_\beta}) \right] \;. \nonumber 
\end{eqnarray}
Applying Eq.~(143) to the cluster $\alpha = K_c(0^-)$ ($S_\alpha = 0$, $S_\beta = \frac{1}{2}$, 
$S=\frac{1}{2}$) (and canceling the phase factor when coupling to different total angular momentum
$J=\frac{1}{2},\frac{3}{2}$ states), we obtain:
\begin{eqnarray}
\gamma_{S}^{\alpha \beta}(\frac{1}{2}|0,\frac{1}{2}) &=& \gamma_L^{K_c \xi_c} 
\;=\;  
\left(\frac{M_{K_c \xi_c}}{M_{\xi_c}}\right) \; \gamma_{\xi_c} \; 
\end{eqnarray}
which is the same result as the classical approximation Eq.~(132). A difference shows up when we
apply Eq.(143) to the cluster $\alpha = K^*_c(1^-)$ ($S_\alpha = 1$, $S_\beta = \frac{1}{2}$,
$S=\frac{1}{2}$):
\begin{eqnarray}
\gamma_{S}^{\alpha \beta}(\frac{1}{2}|1,\frac{1}{2}) &=& \gamma_L^{K^*_c \xi_c} 
\;=\; 
M_{K^*_c \xi_c} \left( \frac{2}{3} \frac{\gamma_{K^*_c}}{M_{K^*_c}} \;-\; 
\frac{1}{3} \frac{\gamma_{\xi_c}}{M_{\xi_c}}\right)\;,
\end{eqnarray}
Inverting as before, we obtain the {\it quantum and CMC consistent} 
gyromagnetic ratio parameter relations:
\begin{eqnarray} 
\gamma_{\xi_c} &=& 
\left(\frac{M_{\xi_c}+M_{K_c}}{M_{K_c}}\right) \; \gamma_L^{K_c \xi_c}  \\
\gamma_{K^*_c} &=& 
\frac{3}{2} \;\left( \frac{M_{K^*_c}+M_{\xi_c}}{M_{\xi_c}}\right) \gamma_L^{K^*_c \xi_c} \;+\;
\frac{1}{2} \;\frac{M_{K^*_c}}{M_{K_c}} 
\left(\frac{M_{K_c}+M_{\xi_c}}{M_{\xi_c}}\right)\gamma_L^{K_c \xi_c}  \\
\gamma_{K_c} &=& 0 \;.
\end{eqnarray}

	\subsection{Color Electro-Dynamics (CED) QMM Hamiltonian}

We now relate the color gyromagnetic ratios to the $A_{\alpha \beta}$ and $B_{\alpha \beta}$ coefficients
in our QMM Hamiltonian Eq.~(95). In the spirit of our CEC and CMC conditions, we treat the form of the
color octet cluster spatial and spin dependent interactions assuming that the primary non-Abelian features
of QCD are contained in the confinement term, leaving the spin-spin and spin-orbit terms in the same form
as the usual electrodynamics electric and magnetic interactions. First consider the spin-orbit interaction
of electrodynamics for a charged boson $\alpha$ and fermion $\beta$. Each cluster separately contributes
to the total spin orbit interaction:
\begin{eqnarray}
H_{so}^{\alpha \beta} &=& H_{so}^{\alpha} \;+\; H_{so}^{\beta} \\
H_{so}^{\alpha} &=& (\frac{1}{2})\; \frac{1}{M_{\alpha \beta}}
\left(\frac{\gamma_{\alpha}}{2 M_{\alpha}}\right) \; \left(\frac{1}{r} \frac{dV}{dr}\right)\;
\vec{S}_\alpha \cdot \vec{L} \\
H_{so}^{\beta} &=& (\frac{1}{2})\; \frac{1}{M_{\alpha \beta}}
\left(\frac{\gamma_{\beta}}{M_{\beta}}\right) 
\left(\frac{1}{r} \frac{dV}{dr}\right)\; \vec{S}_\beta \cdot \vec{L} \;.
\end{eqnarray}
As already noted, the magnetic moment of a fermion is twice as large as a corresponding boson. The factor
of $\frac{1}{2}$ in each term is due to the Thomas precession effect. The spin-orbit interaction is derived
from the energy of a magnetic dipole $\vec{\mu}$ in a magnetic field $\vec{B}$: 
\begin{eqnarray}
H_{so} &=& - \vec{\mu} \cdot \vec{B} \\
\vec{\mu} &=& (\frac{g}{M}) \vec{S} \\
\vec{B} &=& (\frac{g}{M_{\alpha \beta}}) \left( \frac{1}{r} \frac{dV}{dr} \right) \; \vec{L} \;.
\end{eqnarray}
Here $g$ is the fundamental charge (electric or color), $M_{\alpha \beta}$ is the reduced mass of the two
body system, and for a Coulomb potential the derivative term generates a $r^{-3}$ radial dependence. We
write the derivative term explicitly so that the QMM color generalization for the radial dependence can
be modified by the confining term in the potential Eq.~(97). Using S.H.O., confinement generates a constant
term (the classical ``spring constant'') in Eq.~(98), but the extra term would involve a $r^{-1}$ term for
exact linear confinement. We define the effective spin-orbit interaction in terms of a coefficient
$B_{\alpha \beta}$:
\begin{equation}
\left(\frac{B_{\alpha \beta}}{M_{\alpha} M_{\beta}}\right) \; \vec{S} \;=\;
\left( \frac{1}{2 M_{\alpha \beta}} \right) \;
\left[ \left(\frac{\gamma_\alpha}{2 M_\alpha}\right) \; \vec{S}_\alpha \;+\;
\left(\frac{\gamma_\beta}{M_\beta}\right) \; \vec{S}_\beta \right] \;,
\end{equation} 
then the total spin-orbit interaction takes the form expressed in the QMM Hamiltonian Eq.~(95). Matching
coefficients, we define two constants $(a,b)$ which depend on the gyromagnetic ratios
$(\gamma_\alpha,\gamma_\beta)$ and mass ratios:
\begin{eqnarray}
B_{\alpha \beta} \; \vec{S} &=& a\; \vec{S}_{\alpha} \;+\; b \;\vec{S}_{\beta} \\
a &=& \left(\frac{M_\alpha+M_\beta}{2 M_\alpha}\right) \; \left(\frac{\gamma_{\alpha}}{2}\right) \\
b &=& \left(\frac{M_\alpha+M_\beta}{2 M_\beta}\right)
\gamma_{\beta} \;.
\end{eqnarray}
Projecting both sides of Eq.~(156) with the total spin operator $\vec{S}$, and evaluating the spin
expectation values of Eqs.~(136-139), we find the spin-orbit Hamiltonian coefficient:
\begin{equation}
B_{\alpha \beta} \;=\;  \frac{1}{2} \left[ (a+b) + 
     		    (a-b)\left( \frac{S_{\alpha}(S_\alpha+1) - 
     S_\beta(S_\beta+1) }{S(S+1)} \right) \right]  \;.
\end{equation}
Note the $B_{\alpha \beta}$ coefficient depends on the spin quantum numbers $(S,S_\alpha,S_\beta)$ 
of the molecular system. 	

Next consider the more complicated spin-spin $(SS)$ interaction. The form of the $SS$ interaction is
different for the $L=0$ and $L\neq 0$ cases. For completeness, we consider both terms. Our QMM is
restricted to $L=odd$, but we investigate the $L=0$ Hamiltonian to see how it affects our dynamically
restricted spectrum. The electrodynamics inspired form of the interaction is:
\begin{eqnarray}
H_{ss}^{L=0} &=& \alpha_s \; \left(\frac{8 \pi}{3}\right)\; 
\left(\frac{\gamma_\alpha}{2 M_\alpha}\right) \;
\left(\frac{\gamma_\beta}{M_\beta}\right) \; \delta^3(\vec{r}) \; 
\vec{S}_\alpha \cdot \vec{S}_\beta \; \left( \frac{1}{r} \frac{dV}{dr}\right) \\
H_{ss}^{L\neq 0} &=& 
\left(\frac{\gamma_\alpha}{2 M_\alpha}\right)\; \left(\frac{\gamma_\beta}{M_\beta}\right) \;
\left[ 3 (\vec{S}_\alpha \cdot \hat{r})(\vec{S}_\beta \cdot \hat{r}) \;-\; 
\vec{S}_\alpha \cdot \vec{S}_\beta \right] \left( \frac{1}{r} \frac{dV}{dr}\right) \;.
\end{eqnarray}
In the case of $L=0$, the expectation value of $<r^{-3}>$ in the derivative term diverges. However,
the spin-spin term gets a $\delta$-function correction which regulates the divergence, hence the
$<r^{-3}>_{L=0}$ factor is replaced by~\cite{Griffiths87}:
\begin{equation}~\label{eq:wavefunction}
<r^{-3}>_{L=0} \rightarrow \frac{4\pi}{3}\Big |R _{00}(r=0) \Big |^2 \;,
\end{equation}
where $R_{00}(r)$ is the normalized radial wavefunction (in the $n=0,L=0$ state) described in the
next section. In the case of $L\neq 0$, we must take the expectation value of the spin operator in
Eq.~(162).  To simplify the evaluation (and recover the form of the QMM Hamiltonian) we note that to
preserve the CMC constraint with non-variable gyromagnetic ratio parameters, the spin and orbital
angular momentum must couple in maximally aligned or anti-aligned configurations (otherwise, the
dipole color magnetic fields of the quark cluster will not exactly cancel the orbital dipole field). 
Because the CMC condition locks the physical color singlet molecule into maximally aligned
(anti-aligned) configurations,  all quantum substates with zero angular momentum projections
(i.e., $M_L=0,M_{S_\alpha}=0,M_{S_\beta}=0,M_S=0$) are prohibited in our QMM. This physical restriction
(required for consistency with CMC) greatly simplifies the spin-spin operator, since the expectation
value of any $S_x$ or $S_y$ spin projection will be identically zero:
\begin{equation}
\left[ 3 (\vec{S}_\alpha \cdot \hat{r})(\vec{S}_\beta \cdot \hat{r}) \;-\; 
\vec{S}_\alpha \cdot \vec{S}_\beta \right] \;\longrightarrow\; 
2 \; \vec{S}_\alpha \cdot \vec{S}_\beta \;.
\end{equation}
So we arrive at the QMM spin-spin Hamiltonian term expressed in Eq.~(95). The $A_{\alpha \beta}$ is
therefore a simple product of the color gyromagnetic ratios for the $\alpha$ and $\beta$ quark clusters:
\begin{eqnarray}
A_{\alpha \beta} &=& \gamma_\alpha \gamma_\beta \;.
\end{eqnarray}
We now discuss the QMM spatial wavefunctions and resulting mass formula for the molecular pentaquark
exotics.

	\subsection{QMM Spatial Wavefunctions and Expectation Values}

To determine the molecular exotic baryon spectrum, we calculate the first order energy corrections to
the exact energy eigenvalues associated with the confining harmonic oscillator Hamiltonian $\hat{H}_0$:   
\begin{equation}
\hat{H}_0 = - \frac{\nabla^2}{2 \mu} + \frac{1}{2} \mu \omega^2 r^2 \;,
\end{equation}
where in this subsection $\mu$ is the reduced mass
($\mu = M_{\alpha \beta} = M_\alpha M_\beta/(M_\alpha+ M_\beta)$), and $\omega$ is the quantized oscillator
frequency. In this work, we use natural units where $\hbar = c = 1$. The angular eigenfunctions of
$\hat{H}_0$ are the spherical harmonics, $Y_{l m}(\theta,\phi)$. The energy eigenvalues of the unperturbed
oscillator are obtained from the radial Schr\"odinger equation which includes the centrifugal barrier term:
\begin{equation}
\chi'' + \left[ k^2 - \lambda^2 r^2 - \frac{l(l+1)}{r^2} \right] \chi = 0 \;,
\label{eq:diffeq_chi}
\end{equation}
where 
\begin{equation}
k = \sqrt{2 \mu E} \; , \;\;\;\; {\rm and} \;\;\;\; \lambda = \mu \omega \; ,
\end{equation}   
$\chi(r)$ = $r R(r)$ is the reduced radial wave function. Following the usual procedure, the
small/large-$r$ asymptotic behavior of Eq.~(\ref{eq:diffeq_chi}) determines the form of the general solution:
\begin{equation}
\chi(r) = r^{l+1} e^{-\frac{\lambda}{2} r^2} u(r) \;,
\end{equation}
and the function $u(r)$ satisfies the differential equation:
\begin{equation}
\xi \; u''(x) + (l + \frac{3}{2} - \xi) u'(x) + 
\frac{1}{2}(l + \frac{3}{2} - E/\omega) u(x) = 0 \;,
\label{eq:diffeq_u}
\end{equation}
expressed in terms of the dimensionless radial variable $x = \lambda r^2$. The regular solution of
Eq.~(\ref{eq:diffeq_u}) is a confluent (or degenerate) hyper-geometric function
$\Phi[a,b;x]$~\cite{Goldman,Ryzhik}:
\begin{eqnarray}
\Phi[a,b;x] &=& \sum_{j = 0}^{\infty} 
\frac{ \Gamma(a+j)\;\Gamma(b) }{ \Gamma(a) \; \Gamma(b+j) \; \Gamma(1+j) }
	\; x^j \\
& = & 1 + \frac{a}{b} x + \frac{a(a+1)}{b(b+1)} \frac{x^2}{2!} + 
\frac{a(a+1)(a+2)}{b(b+1)(b+2)} \frac{x^3}{3!} + ... \label{eq:phi_seri}\\
u(x) &=& \Phi[\frac{1}{2}(l+\frac{3}{2}-E/\omega),\; l + \frac{3}{2}; \; x] \; \cdot
\end{eqnarray} 
Since physical solutions must have normalizable wave functions, the infinite series in
Eq.~(\ref{eq:phi_seri}) must be truncated after a finite number of terms $a = - n$ (integer $n$), and
hence the oscillator energy levels are quantized: 
\begin{equation}
\frac{1}{2}(l + \frac{3}{2} - E/\omega) = - n \;\;\;\;\; (n = 0,1,2,3,...) \; \cdot
\label{eq:energy_level}
\end{equation}
Solving Eq.~(\ref{eq:energy_level}) we find the harmonic oscillator energy eigenvalues:
\begin{equation}\label{eq:enl}
E_{n l} = (2n + l + \frac{3}{2}) \; \omega  \; ,
\end{equation}
associated with the eigenfunctions:
\begin{equation}
\Psi_{nlm} = N_{n l} \; r^l \; e^{- \frac{\lambda}{2} r^2} \; 
\Phi[-n, \; l+\frac{3}{2}; \; \lambda r^2] \; Y_{lm}(\theta,\phi) \; \cdot
\end{equation}
With these basis functions, we compute the expectation values of the full Hamiltonian to determine the
molecular exotic mass spectrum. In this work, we consider only the states with $n = 0$. Orbital excitations
are prohibited by the CMC constraint, however radial excitation may be permissible. We do not explore the
radial excitations and their consistency with CMC in this work. Since $\Phi[0,b;x] = 1$, the normalized
wave functions are:
\begin{equation}
\Psi_{0lm} = R_{nl}(r)Y_{lm}(\theta,\phi) =
	\left(\frac{2^{l+2}}{(2l+1)!!} \lambda^{l+1} 
\sqrt{\frac{\lambda}{\pi}} \right)^{\frac{1}{2}} \; r^l \; e^{- \frac{\lambda}{2} r^2}
\;  Y_{lm}(\theta,\phi) \; \cdot
\end{equation}
Using these normalized wave functions, a general expression for the expectation value of the radial coordinate
raised to an arbitrary power ($p$) can be evaluated:
\begin{eqnarray}
<  r^p >_{l} &=& \oint d\Omega \int_{0}^{\infty} r^{2+p} \; dr
	\; |\Psi_{0lm}|^2 \\
&=& \left(\frac{2^{l+2}}{(2l+1)!!} \lambda^{l+1} 
\sqrt{\frac{\lambda}{\pi}} \right) \; \Gamma(l + \frac{(3+p)}{2})
	\; \frac{1}{2} \; 
\left(\frac{1}{\lambda}\right)^{l + \frac{(3+p)}{2}} \; \cdot
\end{eqnarray}
Specific expectation values for $l = 0,1$ and $p = 2,1,-1,-2,-3$ are evaluated in Table~\ref{table:oscillator}.

The mass formula for the exotic molecular systems associated with the Hamiltonian of Eq.~(95) is therefore:
\begin{eqnarray}\label{eq:exotic_mass}
M_X^{\alpha \beta} &=& M_\alpha + M_\beta \;+\; (2n + l + \frac{3}{2})\;\omega
	\;+\; C_F^8 \; \alpha_s \; <r^{-1}>  \\
	&+& \left(\frac{A_{\alpha \beta}}{M_\alpha M_\beta}\right) \; 
<{\vec{S}_\alpha\cdot \vec{S}_\beta}> \; [-C_F^8 <r^{-3}> \;+\; \omega^2 \mu \;] \nonumber \\
	&+& \left(\frac{B_{\alpha \beta}}{M_\alpha M_\beta}\right) \;
<{\vec{L}\cdot \vec{S}}>\;[-C_F^8 <r^{-3}> \;+\; \omega^2 \mu \;] \;. \nonumber 
\end{eqnarray}
with $s_\alpha = 0,1$ and $s_\beta = \frac{1}{2},\frac{3}{2}$. The oscillator frequency $\omega$ in
Eq.~(\ref{eq:exotic_mass}) is given by the mass splitting between the proton and its first radial excitation
using Eq.~(175):
\begin{equation}
\omega \; = \; \frac{1}{2} [M_{N^*(1440)} - M_p(938)] = 0.251 \; ; \\
\end{equation}

\begin{figure}[!hbt]
\begin{center}
\mbox{
\epsfxsize = 8cm
\epsfysize = 8cm
\epsffile{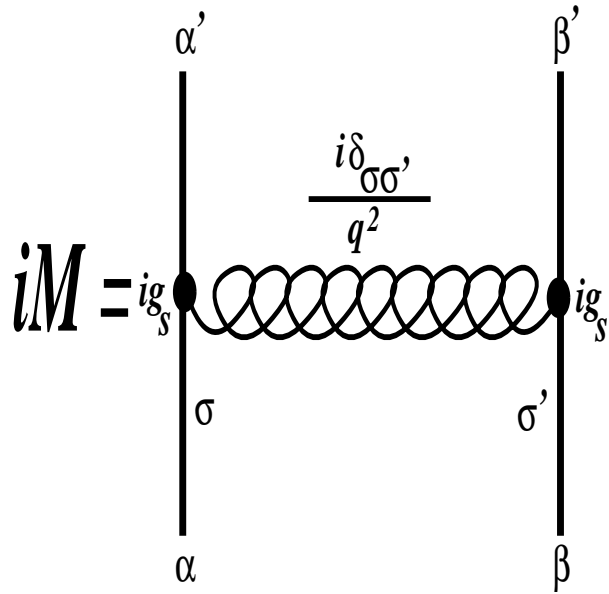}
}
\end{center}
	\caption{Feynman diagram of the one gluon exchange between the the color octet clusters.}
	\label{fig:colorfactor}
\end{figure}

The usual color factor between color triplet quarks forming a color singlet 
$C_F(3 \otimes \bar{3} \rightarrow 1) = -4/3$ is not appropriate for one-gluon exchange between color octet
clusters. Therefore, we will derive the correct factor $C_F^8 = C_F(8 \times 8 \rightarrow 1)$ for octet
bonded clusters. Evaluating the Feynman diagram for gluon exchange between color octet (point-like) particles
(Fig.~\ref{fig:colorfactor}), we have:
\begin{eqnarray}
C_F^8 &=& (\frac{1}{\sqrt{8}})^2 \; (i)^2  \; 
f^{\alpha \alpha' \sigma} \; f^{\beta \beta' \sigma'} \; 
\delta_{\sigma \sigma'} \; \delta_{\alpha \beta}  \; 
\delta_{\alpha' \beta'} \\
&=& -(1/8)(6) [ \; (f^{123})^2 \;+\; 6 \;(f^{147})^2 \;+\; 2 \;(f^{458})^2 \;]
\rightarrow - 3   \;,
\end{eqnarray}
which involves a trace over the product of $SU(3)$ structure constants $f^{\alpha \beta \gamma}$, and
projecting octet color functions $(a^j_\gamma)$ onto singlet initial $|1>$ and final $|1'>$ state color functions:
\begin{eqnarray}
&& |1> \;=\; \left(a_\alpha^1 a_{\beta}^2 \right)_1 \;=\; 
\frac{1}{\sqrt{8}} \; \delta_{\alpha \beta} \\
&& |1'> \;=\; \left(a_{\alpha'}^3 a_{\beta'}^4\right)_1 \;=\; 
\frac{1}{\sqrt{8}} \; \delta_{\alpha' \beta'} \;.
\end{eqnarray}
Note the color octet-octet interaction generates a strongly attractive one gluon exchange potential,
$V_{1g}(r) = C_F^8 \alpha_s/r$.

The last step in constructing the QMM spatial wavefunctions involves coupling cluster spins
$(S_\alpha,S_\beta)$, total spin $S$, and orbital angular momentum $L$ to overall angular momentum $J$. The
angular momentum $J$-states are calculated using the superposition of $|L>|S>$ basis states with CMC restricted 
Clebsch-Gordon coupling factors (all magnetic substates equal to zero are prohibited in the CMC restricted sum):
\begin{eqnarray}
& &|\Psi_{\alpha,\beta}(r,\theta,\phi)||S,L,M_L,J,M_J> \;=\;
\sum_{M_{S_\alpha},M_{S_\beta},M_S}^{CMC}  \\ 
&& C(J,M_J|L,M_L,S,M_S) \; \nonumber 
C(S,M_S|S_\alpha,M_{S_\alpha},S_\beta,M_{S_\beta}) \; 
R_{NL}(r)\;  Y_L^{M_L}(\theta,\phi)\;.  \nonumber
\end{eqnarray}
Since the CMC principle restricts the angular momentum substates, the standard Clebsch-Gordon factors are
renormalized in the QMM sum over states (details omitted). We now discuss our QMM parameter assignments and the
resulting physical spectrum of pentaquark molecules.

	\subsection{QMM Parameters and Spectrum Results}

The QMM cluster masses are not a priori known (since the color octet clusters can not be measured in unbound
isolation). However, their masses should be comparable to the physical $K, K^*$ and $\Lambda$ masses due to the
dominating contributions from constituent quark masses. Since the quarks inside each cluster combine to form a
color octet, we anticipate a mass shift (increase) originating from the repulsive color factor of one gluon exchange
between quarks in the octet color configuration. To estimate this mass shift, we calculate the color factor of
one gluon  exchange between $q \bar{q}$ coupled to a color octet,
$C_F(3 \otimes \bar{3}\rightarrow 8) = + \frac{1}{6}$, whereas the color factor for the color singlet is
$C_F(3 \otimes \bar{3} \rightarrow 1) = -\frac{4}{3}$. So the estimated mass shift due to one gluon exchange
between quarks inside a color octet cluster is: $\delta M = \frac{3}{2} \alpha_s <\frac{1}{r}>$. 

For a typical hadronic size ($<r> \approx 0.5$~fm), we deduce the mass shift $\delta M \approx 0.1$~GeV. Since
this is a small number compared to the bare cluster masses (i.e., physical $K,K^*$ and $\Lambda$ masses), 
and because the $\xi$-mass is an unknown adjustable constant which is fixed by the lightest pentaquark candidate,
we simply use the physical $K,K^*$ masses for the $\alpha(q \bar{s})$ clusters and absorb the mass shift ambiguity
into the unknown phenomenological $\xi$-mass, which is the  only unconstrained parameter of QMM
(Table~\ref{table:exotic_theory_parameters}). The calculated QMM masses of the exotic (pentaquark) states are
summarized in Table~\ref{table:exotic_theory_result}.



\section{Experimental status}\label{ExperimentalStatus}

In this section, we review the current status on experimental data focusing on associated electromagnetic production
of strangeness systems of protons. In particular, (re-) analysis of experiments were done in the spirit of finding
potential narrow structures to support our QMM calculation.

	\subsection{Photo-production channel}\label{section:photoproduction}

Guided by the recent claim of three narrow resonances from the SPHINX collaboration~\cite{SPHINX95,SPHINX97,SPHINX99},
we re-examined the published data on K-meson photo-production using the $p(\gamma,K)[\Lambda,\Sigma^0]$ reaction
from the SAPHIR detector at ELSA~\cite{saphir98}. 

\begin{figure}[!hbt]
\begin{center}
\mbox{
        \epsfxsize = 12cm
	\epsfysize = 12cm
        \epsffile{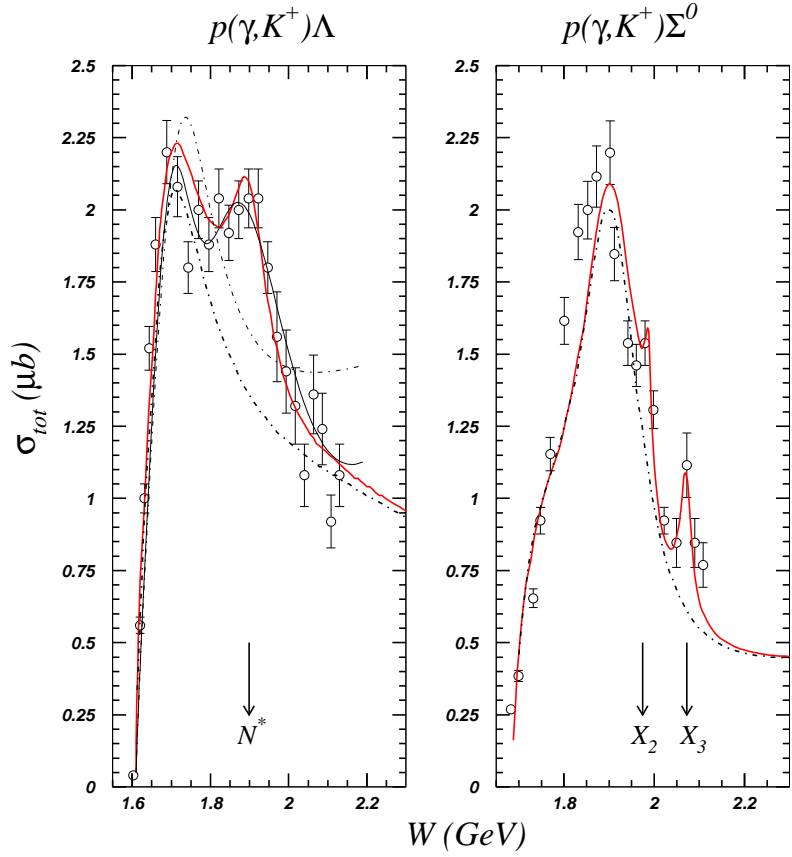}
}
        \caption{$\Lambda$ (left) and $\Sigma ^0$ (right) invariant mass distributions in photo-production
	from~\protect\cite{saphir98} compared to the model prediction of~\protect\cite{GW01,GW02} when the narrow
	baryon X-resonances are taken into account (solid line), and without X (dot-dashed line). The dotted
	line in the $\Lambda$ channel is the prediction from~\protect\cite{Bennhold99}. See text for details.}
        \label{fig:saphir_clas_june99} 
\end{center}
\end{figure}

The invariant mass distributions for both $\Lambda$ and $\Sigma$ hyperons are compared to the background generated by
a well constrained hadrodynamical Lagrangian model prediction of~\cite{GW01,GW02} in Fig.~\ref{fig:saphir_clas_june99}.
The various curves account for when the narrow baryon X-resonances are included (solid line), or not included (dot-dashed
line). We assume assume relativistic Breit-Wigner shapes for the resonances, which are superimposed on the smooth
background. The masses, total widths and signal strengths are adjusted to optimize $\chi ^2$ agreement with the
analyzed data (Table~\ref{table:saphir}). 

Overall, the qualitative agreement between the data and the Lagrangian model prediction is significantly improved for
both hyperon production channels when the X-baryon resonances are included.  A detailed description of the hadronic
Lagrangian model employed is under preparation for submission~\cite{GW01,GW02}.

	\subsection{Electro-production channel}\label{section:electroproduction}

The kaon photo-production results described in the previous section provided motivation to investigate the equivalent
(virtual photon mediated) kaon electro-production reactions. The elementary process for $p(e,e'K)[\Lambda,\Sigma^0]$
is well described at low $Q^2$\cite{Gueye00}. Analysis of recent Jefferson Lab data has shown that isobar based models
provide a good description of electro-production observables~\cite{Niculescu98,Mohring03}. We employ the same isobar
model~\cite{GW01,GW02} to check for consistency with the photo-production results. We analyze dedicated kaon
electro-production data that were taken in December 1999 during the second stage of experiment
E91016~\cite{Zeidman,Reinhold98} at Jefferson Lab in the experimental Hall C using the 100\% duty cycle of the CEBAF
machine at a fixed beam energy of 3.245~GeV. 

The scattered electrons were detected in the High Momentum spectrometer (HMS, $\Delta P/P = \pm 10\%$,
$\delta P/P = 10^{-3}$, $\Delta\Omega = 6.7$~msr), set at 19.64$^\circ$ and with a central momentum of 1.432~GeV/c.
The K-mesons were detected in the Short Orbit spectrometer (SOS, $\Delta P/P = \pm 20\%$, $\delta P/P = 10^{-3}$,
$\Delta\Omega = 7.7$~msr) along the virtual photon direction at 14.23$^\circ$ with a central momentum of 1.290~GeV/c.
The corresponding invariant mass and $Q^2$ settings were 1.925~GeV and 0.55~(GeV/c)$^2$, respectively. The detector
packages in both spectrometers are similar and include: two planes of drift chambers (used for tracking information),
four planes of plastic scintillators (used for trigger formation, and time-of-flight (TOF) information), a gas
\v{C}erenkov (for $e/\pi$ separation), and a lead glass shower counter (for electron identification). The SOS had two
additional \v{C}erenkov counters to improve identification of the kaons: an aerogel \v{C}erenkov (for $\pi^+/K^+$
separation), and a lucite \v{C}erenkov (for $p/K^+$ separation). The detector performances achieved were similar to
the ones obtained in earlier kaon experiments performed in this experimental hall~\cite{Niculescu98,Mohring03}.

In the hadron arm, the $K^+$ were identified using the aerogel, the lucite and TOF. In the electron arm, we used the
gas \v{C}erenkov and Pb-glass. A sample of 8262 coincident kaons were then reconstructed which includes subtraction of
the background events from the target cell walls (0.6\%) and random coincident events (real-to-random ratio about 10/1).

\begin{figure}[!hbt]
\begin{center}
\mbox{
        \epsfxsize = 10cm
        \epsfysize = 10cm
        \epsffile{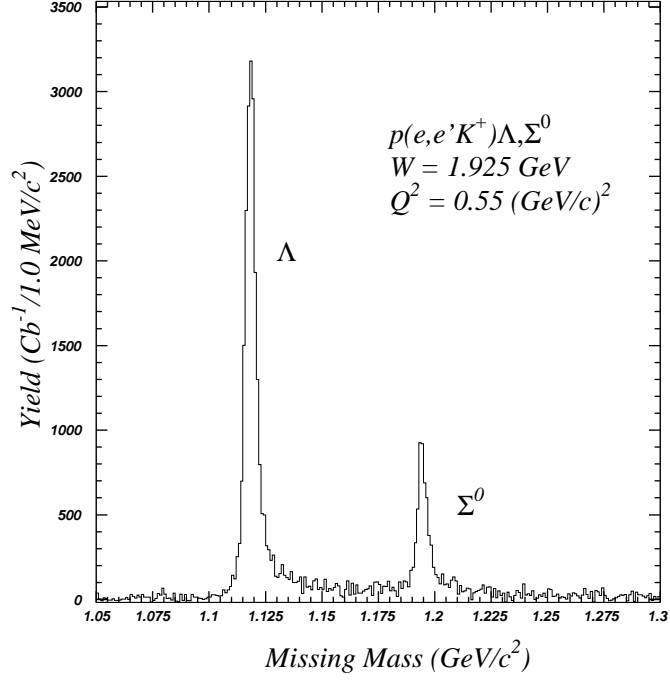}
}
        \caption{Missing mass distribution for the kaon electro-production reaction~\protect\cite{Zeidman,Reinhold98}.
	Random coincidences and target wall contributions are subtracted.}
        \label{exotic_mm_13dec99}
\end{center}
\end{figure}

The two hyperons produced during the reaction can be seen at their expected location (Fig.~\ref{exotic_mm_13dec99}):
$M_{\Lambda(1116)} = (1116.40 \pm 0.66)$~MeV [$\Gamma_{\Lambda} = (4.82 \pm 0.65)$~MeV] and
$M_{\Sigma ^0(1193)} = (1192.7 \pm 0.12)$~MeV [$\Gamma_{\Sigma} = (4.86 \pm 1.60)$~MeV].

\begin{figure}
\begin{center}
\mbox{
	\hspace{-0.5cm}
        \epsfxsize = 10cm
        \epsfysize = 10cm
        \epsffile{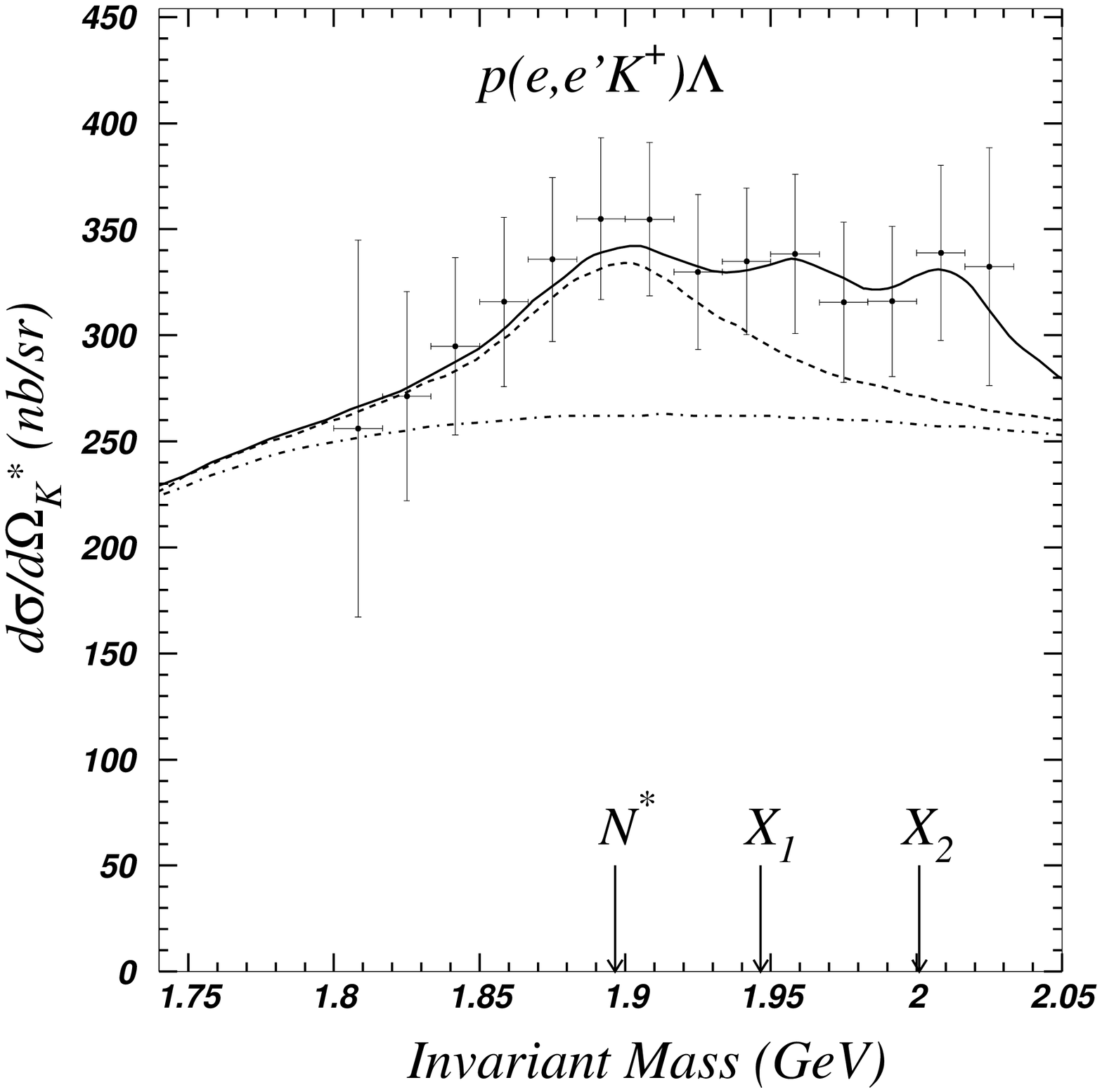}
	\hspace{-0.5cm}
        \epsfxsize = 10cm
        \epsfysize = 10cm
        \epsffile{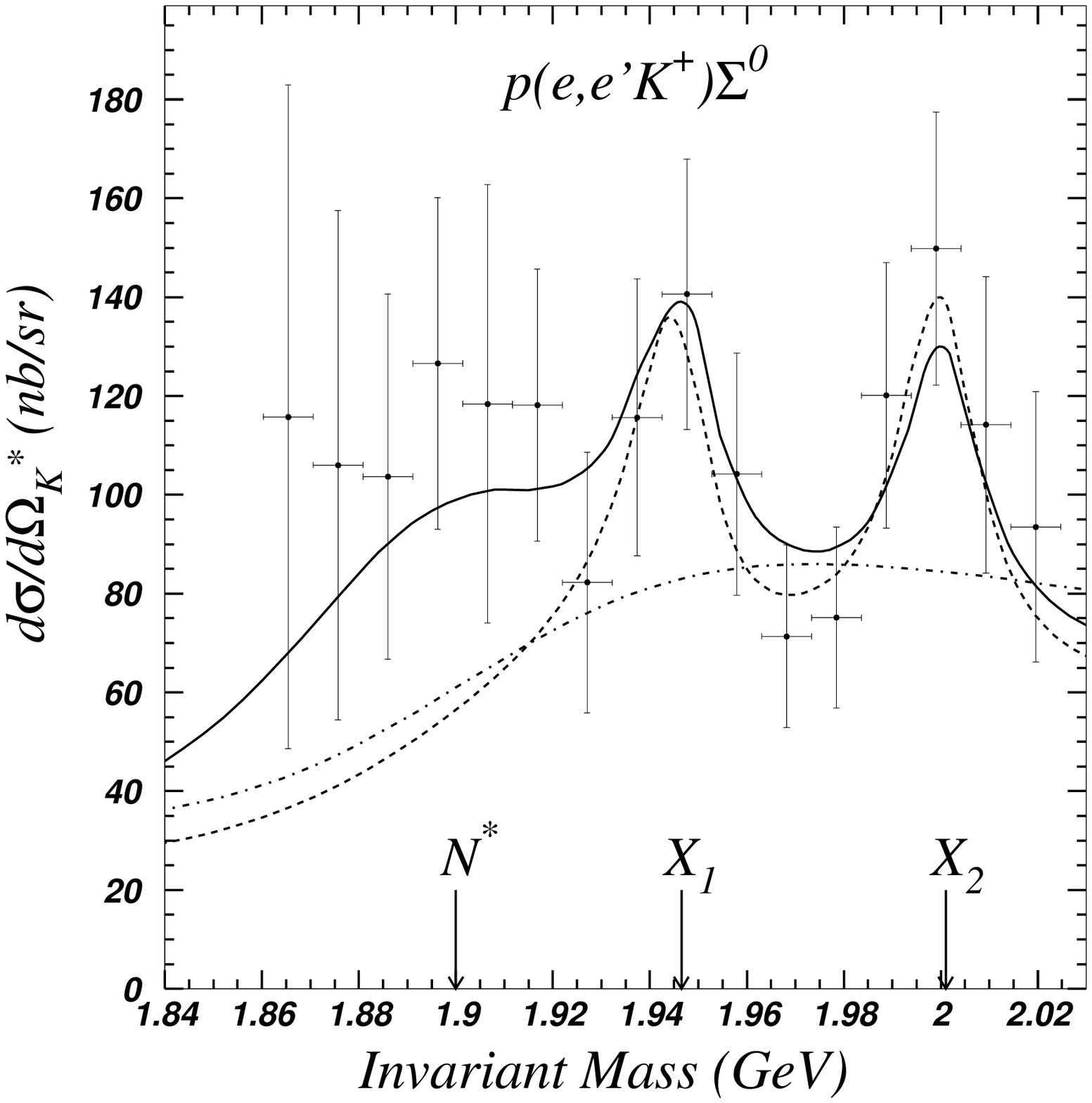}
}
        \caption{Invariant mass distributions of the $\Lambda$ (left) and $\Sigma ^0$ (right) hyperons. The model
	prediction is compared to the experimental data~\protect\cite{Zeidman,Reinhold98} without (dot-dashed line)
	and with (dashed and solid line) coupling to the narrow X-resonances using the prediction
	of~\protect\cite{GW01,GW02}. See text for details.}
        \label{exotic_exp_wjc}
\end{center}
\end{figure}

The invariant mass distributions are shown in Fig.~\ref{exotic_exp_wjc} in addition to our model differential cross
section of~\cite{GW01,GW02} for the $\Lambda$ (left) and $\Sigma^0$ (right) hyperons. The $\chi ^2$ fits of the three
enhancements using relativistic Breit-Wigner shapes are listed in Table~\ref{table:JLab99}.

As a result of the different bin sizes used in the analysis of the data in~\cite{saphir98} (30~MeV)
and~\cite{Zeidman,Reinhold98} ($\sim$12.55~MeV), the new Jefferson Lab electro-production invariant mass distributions
are able to resolve two nearly degenerate resonances, $X_{1}$ and $X_{2}$, in the vicinity of the SAPHIR
$X_{2}$~\cite{saphir98}. We believe that due to the large bining of the SAPHIR data, the first peak in $\Sigma ^0$
production at 1.974~GeV averages over the JLab $X_{1}$ and $X_{2}$, which contribute in the same SAPHIR energy bin. The
resulting averaged value, $M_{\overline{X_{1}\oplus X_{2}}} = (1974.1 \pm 6.7)$~MeV
[$\Gamma _{\overline{X_{1}\oplus X_{2}}} = (97.7 \pm 55.1)$~MeV], is less than 1-$\sigma$ of the extracted SAPHIR $X_2$
mass value, and the decay width is approximately equal to the sum of the two JLab $X_1$ and $X_2$ resonance widths. As
in photo-production, we find optimum agreement between our model prediction and the electro-production data when the
X-resonance degrees of freedom are included~\cite{GW01,GW02}.

	\subsection{Summary}

Although the resonances extracted have fairly poor statistics, there is a clear definite pattern between all of the
experimental data. This gives us a strong confidence for the existence of the narrow states extracted. We summarize the
results or our analysis for all experimental data sets in comparison with the QMM spectrum predictions in
Table~\ref{table:exotic_experiment_QMM}.

Although a dedicated new high precision electroproduction experiment is planned to be performed for confirming and
extracting the properties of these resonances~\cite{GW03}, additional data in photo- and electro-production in that $W$
range will be needed in the near future to shed light on the possible existence of the combined SPHINX, SAPHIR and Hall
C data resonances discussed here. We would like to stress that the $W$ distribution of the differential cross section
would have to be kinematically constraint to center-of-mass angles between the virtual photon and outgoing kaon to be
$\theta _{\gamma^*-K^+} \leq 20-30 ^\circ$~\cite{GW01,GW02}) due to the (expected) strong forward peaking effect of these
states on observable quantities.

\section{Conclusion}

We have developed a quark molecular model (QMM) of $S=0$ pentaquark baryons assuming color-octet bonded
$(q\bar{s})$-$(uds)$ molecular structure, and predict the low lying mass spectrum (4-states) using a color
electrodynamics (CED) Hamiltonian. We have introduced the idea of color magnetic confinement (CMC) analogous to the
well known color electric confinement (CEC) of traditional quark models. The CMC condition limits the allowed physical
states in the QMM spectrum and generates consistency relations between the color gyromagnetic ratio factors of the
QMM-CED Hamiltonian. This novel prediction stands in contrast to the usual outcome of quark model spectroscopy (which
generate an  overabundance of excited states) and provides a possible explanation of the dilemma  of ``{\it missing
resonances}''.

To show the plausibility of our pentaquark spectrum predictions, we have reported experimental evidence of four narrow
structures in the vicinity of $W = 2$~GeV in associated strangeness production reactions using the recent ELSA data in
kaon photo-production~\cite{saphir98} and Jefferson Lab data in kaon electro-production~\cite{Zeidman,Reinhold98}. One
resonance structure is not consistent with the QMM mass prediction and requires a typical (large $\Gamma \sim 200$ MeV) 
hadronic decay width, and hence is assumed to be a conventional (possibly missing) $N^*$ resonance, the $N^*(1898)$.

The suggested {\it candidate} pentaquark states of this analysis have average masses of 1.812, 1.962~{\footnote{The
SAPHIR~\cite{saphir98} datum at 1974~GeV was not included in this average for consistency with the adopted analysis
method of the data sets from the experiments discussed in this document.}}, 2002 and 2.064~GeV, all of which have been
identified with anomalously narrow decay widths (few tens of MeV). Two of those resonances (1.943~GeV and 2001.7~GeV)
seen in the recent JLab data are believed to appear as an unresolved single resonance in the SAPHIR data at 1.974~GeV
(due to relatively coarse energy bining). The fitted Breit-Wigner resonances align closely with the 4-predicted pentaquark
states of the QMM, with masses: $X_0(1801)[1/2^+]$, $X_1(1967)[3/2^+]$, $X_2(1999)[1/2^+]$, and $X_3(2064)[3/2^+]$.

Additional high resolution experiments are required to establish the unambiguous existence and properties (i.e., masses,
decay widths, branching fractions, hadronic coupling strengths, form factors, etc.) of the candidate $S=0$ pentaquark
baryons suggested in this work. Our results provide a strong indication of possible crypto-exotic pentaquark baryon
resonances in associated strangeness production channels, and support the need for investigating reaction channels other
than non-strange ($\pi, \rho, \omega,$ etc.) production to look for missing quark model resonances~\cite{RobCap9498} 
and QCD allowed exotic states. 

\begin{acknowledgements}
The authors acknowledge the support of the staff of the Accelerator and Physics Divisions of the Thomas Jefferson
National Accelerator Facility. They also would like to express their gratitude to P. Ambrozewicz, K. Baker, R. Ent and M.
Khandaker for valuable discussions. RW appreciates the encouraging spirit and valuable input of Nathan Isgur during the
early development of ideas realized in the QMM. This work was supported in part by the National Science Foundation
(Grant No. NSF-9633750).
\end{acknowledgements}

\newpage

\begin{table}[!hbt]
\begin{center}
\begin{tabular}{c|c|c|c}
$p$	& $< r^p >_l$	& $l=0$	& $l=1$ \\
\hline
\hline
2	& $< r^2 > $	& $\frac{3}{2 \lambda}$
			& $\frac{5}{2\lambda}$ \\
1	& $< r > $	& $\frac{2}{\lambda} \sqrt{\frac{\lambda}{\pi}}$
			& $\frac{8}{3\lambda} \sqrt{\frac{\lambda}{\pi}}$ \\
-1	& $< r^{-1} >$	& $2 \sqrt{\frac{\lambda}{\pi}}$
			& $\frac{4}{3} \sqrt{\frac{\lambda}{\pi}}$ \\
-2	& $< r^{-2} >$	& $ 2 \lambda $
			& $\frac{2}{3} \lambda $ \\
-3	& $< r^{-3} >$	& -
			& $\frac{4 \lambda}{3} \sqrt{\frac{\lambda}{\pi}}$
\end{tabular}
	\caption{Expectation values of the radial coordinate in the
	oscillator basis.}
	\label{table:oscillator}
\end{center}
\end{table}

\begin{table}[!hbt]
\begin{center}
\begin{tabular}{c|c|c|c}
Cluster ($\alpha,\beta$) & {Cluster Mass}	& {Gyromagnetic Ratio $\gamma$} & 
  {Gyromagnetic Ratio $\gamma$}              \\
&       $[GeV]$                &  (Quantum Set)    & (Classical Set) \\
\hline
\hline
$\; \alpha = K_c(0^-|q\bar{s})\; $      & $M_{K_c} = M_K$ = 0.497		& 0		& 0	\\
$\alpha = K^*_c(1^-|q\bar{s})$     	& $\; M_{K_c^*} = M_{K^*} = 0.896 \;$	& 2.189     	& 2.462 \\
$\beta = \xi_c(\frac{1}{2}^+|uds)$     	& $M_{\xi_c}$ = {\bf 0.982}	        & 1.648		& 1.648		
\end{tabular}
	\caption{QMM parameters. The cluster color gyromagnetic ratios are  fixed by the CMC constraint in
	terms of the cluster masses [$GeV$]. We fix the color octet cluster masses to equal the corresponding
	flavor equivalent $K$ and $K^*$ masses. The only free parameter is the flavor singlet color octet mass
	$\xi_c$ (in bold type) adjusted to reproduce the lightest pentaquark candidate ($X_0(1800)$) observed
	in the SPHINX data~\protect\cite{SPHINX95,SPHINX97,SPHINX99}.}
	\label{table:exotic_theory_parameters}
\end{center}
\end{table}

\begin{table}[!hbt]
\begin{center}
\begin{tabular}{c||c||c|c}
& & {QMM Mass [GeV]}	& {QMM Mass [GeV]} \\
Name	&   $X_{\alpha \beta}$$(J^\pi|L,S)$ & {Quantum Parameters}  & {Classical Parameters} \\
\hline
\hline
$\;X_0(\frac{1}{2}^+)\;$ & $\;X_{K_c \xi_c}(\frac{1}{2}^+|1,\frac{1}{2})\; $
		     	& 1.801			& 1.801			\\
$X_1(\frac{3}{2}^+)$ 	& $X_{K_c \xi_c}(\frac{3}{2}^+|1,\frac{1}{2}) $
		     	& 1.967			& 1.967			\\
$X_2(\frac{1}{2}^+)$ 	& $X_{K^{*}_{c} \xi_c}(\frac{1}{2}^+|1,\frac{1}{2})$ 
			& 1.999			& 1.972			\\
$X_3(\frac{3}{2}^+)$ 	& $ X_{K^{*}_{c} \xi_c}(\frac{3}{2}^+|1,\frac{1}{2}) $
			& 2.064			& 2.049			
\end{tabular}
	\caption{The comparison between calculated QMM exotic masses [$GeV$] 
	using quantum consistent and classical parameters for the gyromagnetic ratios. Note the 
	different parameter sets only affect the $K_c^*$ gyromagnetic ratio, so the two lightest 
	exotics with ($K_c$-$\xi_c$) substructure are unaffected. The S.H.O. oscillator frequency 
	was fixed by the $N^*(1440)-N(938)$ mass splitting: $\omega = 0.251$ GeV.}
	\label{table:exotic_theory_result}
\end{center}
\end{table}

\begin{table}
\begin{center}
\begin{tabular}{l||c|c}
SAPHIR Data Fit	& $\; p(\gamma,K^+)\Lambda \;$ & $\; p(\gamma,K^+)\Sigma^0 \;$	\\
\hline
$M_{X_0}$ [MeV]		& N/S				& N/S			\\
$\Gamma_{X_0}$ [MeV]	& N/S				& N/S			\\
$C_{X_0}$		& N/S				& N/S			\\
C.L.			& -				& -			\\
\hline
$M_{N^*}$ [MeV]		& $\; 1898.4 \pm 9.1 \; $	& N/S			\\
$\Gamma_{N^*}$ [MeV]	& $246.2 \pm 57.0$		& N/S			\\
$C_{N^*}$		& $1.11 \pm 0.04$		& N/S			\\
C.L.			& 93.42\%			&	-		\\
\hline
$M_{X_1}$ [MeV]		& N/S 				& N/S			\\
$\Gamma_{X_1}$ [MeV] 	& N/S 				& N/S			\\
$C_{X_1}$		& N/S				& N/S			\\
C.L.			& -				& -			\\
\hline
$M_{X_{2}}$ [MeV]	& N/S 				& $\; 1974.2 \pm 3.2 \;$\\
$\Gamma_{X_{2}}$ [MeV] 	& N/S 				& $114.5 \pm 11.3$	\\
$C_{X_{2}}$		& N/S				& $1.10 \pm 0.04$	\\
C.L.			& -				& $> 99\%$		\\
\hline
$M_{X_3}$ [MeV]		& N/F				& $2072.7 \pm 4.7$	\\
$\Gamma_{X_3}$ [MeV]	& N/F				& $113.5 \pm 30.9$	\\
$C_{X_3}$		& N/F				& $1.27 \pm 0.08$	\\
C.L.			& -				& 73.58\%		\\
\end{tabular}
	\caption{Masses and widths which characterize the narrow resonances fits to the
	$p(\gamma,K)[\Lambda,\Sigma ^0]$ SAPHIR data~\protect\cite{saphir98}. $M_{X_i}$ are the masses,
	$\Gamma _{X_i}$ the widths, and $C_{X_i}$ the signal strengths. N/S: not seen in this channel.
	N/F: peak not fitted. C.L.: confidence level. See text for details.}
	\label{table:saphir}
\end{center}
\end{table}

\begin{table}[!hbt]
\begin{center}
\begin{tabular}{l||c|c}
JLab Data Fit $\;$ & $p(e,e'K^+)\Lambda$	& $\; p(e,e'K^+)\Sigma^0 \;$	\\
\hline
$M_{X_{0}} [MeV]$	& N/F				&  O/A  		\\
$\Gamma_{X_{0}}$ [MeV] 	& N/F				& O/A 			\\
$C_{X_{0}}$		& N/F    			& O/A			\\
C.L.			& -				& -			\\
\hline
$M_{N^*}$ [MeV]		& $1896.5 \pm 16.4$ 		& N/F			\\
$\Gamma_{N^*}$ [MeV] 	& $\;  239.09 \pm 121.4 \;$ 	& N/F			\\
$C_{N^*}$		& $1.18 \pm 0.01$		& N/F			\\
C.L.			& $> 99\%$			& 	-		\\
\hline
$M_{X_{1}} [MeV]$	& N/F				& $\; 1946.5 \pm 4.2 \;$ \\
$\Gamma_{X_{1}}$ [MeV] 	& N/F				& $43.6 \pm 16.0$	\\
$C_{X_{1}}$		& N/F				& $1.53 \pm 0.26$	\\
C.L.			& -				& $> 99\%$		\\
\hline
$M_{X_{2}}$	[MeV] 	& N/F				& $2001.7 \pm 5.2$	\\
$\Gamma_{X_{2}}$ [MeV] 	& N/F				& $54.1 \pm 23.2$	\\
$C_{X_{2}}$		& N/F				& $1.56 \pm 0.29$	\\
C.L.			& -				& $> 99\%$		\\
\hline
$M_{X_{3}}$	[MeV] 	& O/A				&  O/A 			\\
$\Gamma_{X_{3}}$ [MeV] 	& O/A				& O/A 			\\
$C_{X_{3}}$		& O/A				& O/A			\\
C.L.			& -				& -			\\
\end{tabular}
	\caption{Masses and widths characteristics of the narrow resonances fit to the
	$p(e,e'K)[\Lambda,\Sigma ^0]$ Jefferson Lab (JLab) data~\protect\cite{Zeidman,Reinhold98}.
	N/F: peak not fitted. O/A: out of acceptance. C.L.: confidence level. Note the large width associated
	with the assumed $N^*$ peak (absent from the QMM spectrum).}
	\label{table:JLab99}
\end{center}
\end{table}

\begin{table}[!hbt]
\begin{center}
\begin{tabular}{l||c||cc||cc||c}
		& SPHINX
		& \multicolumn{2}{c||}{SAPHIR}
		& \multicolumn{2}{c||}{JLab}
		& {QMM} \\
\hline
		& $^{12}C(p,K^+)X$
		& $p(\gamma,K^+)\Lambda$ & $p(\gamma,K^+)\Sigma^0$
		& $p(e,e'K^+)\Lambda$ & $p(e,e'K^+)\Sigma^0$ 
		& ($J^\pi$)\\
\hline
\hline
$M_{X_{0}}$	& ($1812.0 \pm 7.0$)
		& N/S			& N/S
		& N/S			& N/S
		& 1801						\\
$\Gamma_{X_{0}}$& ($56 \pm 1$)
		& N/S			& N/S
		& N/S			& N/S
		& $(\frac{1}{2}^+)$				\\
$C_{X_{0}}$	& -
		& N/S			& N/S
		& N/S			& N/S
		&						\\
\hline
$M_{N^*}$	& N/S
		& ($1898.4 \pm 9.1$)	& N/S
		& ($1896.5 \pm 16.4$)	& N/S
		& Missing					\\
$\Gamma_{N^*}$& N/S
		& ($246.2 \pm 57.0$)	& N/S
		& ($239.1 \pm 121.4$)	& N/S
		& ($?^?$)					\\
$C_{N^*}$	& -
		& $1.11 \pm 0.04$	& N/S
		& $1.18 \pm 0.01$	& N/S
		&						\\
\hline
$M_{X_{1}}$	& N/S
		& N/S			& N/S
		& ($1946.5 \pm 4.2$)	& ($1938.4 \pm 6.7$)
		& 1967						\\
$\Gamma_{X_{1}}$& N/S
		& N/S			& N/S
		& ($43.6 \pm 16.0$)	& ($239.9 \pm 57.9$)
		& $(\frac{3}{2}^+)$				\\
$C_{X_{1}}$	& -
		& N/S			& N/S
		& $1.53 \pm 0.26$	& $1.43 \pm 0.03$
		&						\\
\hline
$M_{X_{2}}$	& ($1997.0 \pm 7.0$)
		& N/S			& ($1974.2 \pm 3.2$)
		& O/A			& ($2001.7 \pm 5.2$)
		& 1999						\\
$\Gamma_{X_{2}}$& ($91 \pm 17$)
		& N/S			& ($114.5 \pm 11.3$)
		& O/A			& ($54.1\pm 23.2$)
		& $(\frac{1}{2}^+)$				\\
$C_{X_{2}}$	& -
		& N/S			&$1.10 \pm 0.04$
		& O/A			& $1.56 \pm 0.29$
		&						\\
\hline
$M_{X_3}$	& ($2052.0 \pm 6$)
		& ($2068.6 \pm 4.3$)	& ($2072.7 \pm 4.7$)
		& O/A			& O/A
		& 2064						\\
$\Gamma_{X_3}$& ($35.0 \pm 29$)
		& ($112.1 \pm 24.3$)	& ($113.5 \pm 30.9$)
		& O/A			& O/A
		& $(\frac{3}{2}^+)$				\\
$C_{X_3}$	& -
		& $ 1.36 \pm 0.11$	& $1.27 \pm 0.08$
		& O/A			& O/A
		&						\\
\end{tabular}
	\caption{Experimental masses and widths of the narrow resonances fit to the $^{12}C(p,K^+)X$ SPHINX
	data~\protect\cite{SPHINX95,SPHINX97,SPHINX99}, $p(\gamma,K)[\Lambda, \Sigma ^0]$ SAPHIR
	data~\protect\cite{saphir98}, and $p(e,e'K) [\Lambda,\Sigma ^0]$ Jefferson Lab (JLab)
	data~\protect\cite{Zeidman,Reinhold98}. $M_X$ are the masses [MeV], $\Gamma _X$ the FWHMs [MeV],
	and $C_X$ the Breit-Wigner coupling strengths. N/S: not seen in this channel, O/A: out of acceptance.
	The QMM masses were calculated using the quantum consistent parameter set.}
	\label{table:exotic_experiment_QMM}
\end{center}
\end{table}

\begin{references}
\bibitem{Isgur77} N. Isgur and G. Karl, Phys. Lett. {\bf 72B}, 109
(1977); Phys. Rev. D {\bf 23}, 817 (1981).
\bibitem{GW01} P. Gu\`eye and R. A. Williams, Photo-production of exotic $S = 0$ baryon resonances
	through final state interactions in $K^+\Lambda$ and $K^+\Sigma^0$ reactions, to be submitted
	to Phys. Rev. C (2003). 
\bibitem{GW02} P. Gu\`eye and R. A. Williams, Electro-production of exotic $S = 0$ baryon resonances
	through final state interactions in $K^+\Lambda$ and $K^+\Sigma^0$ reactions, to be submitted
	to Phys. Rev. C (2003). 
\bibitem{SAPHIR03} The Saphir Collaboration, {hep-ex/0307083} - Accepted for publication
in Phys. Lett. {\bf B} (2003).
\bibitem{CLAS03} The CLAS Collaboration, preprint hep-ex/0307018 - Submitted to
Phys. Rev. Lett. (2003).
\bibitem{Lipkin87} H. Lipkin, Phys. Lett., {\bf B195}, 484 (1987).
\bibitem{WeinIsgur90} J. Weinstein and N. Isgur, Phys. Rev. D,
{\bf 41}, 2236 (1990).
\bibitem{Schumacher97} R. A. Schumacher, Phys. Rev. C, {\bf 56},
2774 (1997).
\bibitem{CHS79} M. de Crombrugghe, H. H\"ogaasen and P. Sorba,
Nucl. Phys., {\bf B156}, 347 (1979).
\bibitem{HS78} H. H\"ogaasen and P. Sorba, Nucl. Phys., {\bf B145},
347 (1978).
\bibitem{OZI} S. Okubo, Phys. Lett. {\bf 5}, 165 (1963); Z. Zweig,
CERN Report 8419/TH-412 (1964); T. Iizuka, K. Okada, and O. Shito,
Prog. Theor. Phys. Suppl. {\bf 37}, 38 (1966).
\bibitem{CJKTW74} A. Chodos, R. L. Jaffe, K. Johnson, C. B. Thorn and
V. F. Weisskopf, Phys. Rev. D, {\bf D} (1974).
\bibitem{JT76} K. Johnson and C. Thorn, Nucl. Phys., {\bf B41}, 397
(1978).
\bibitem{Griffiths87} D. Griffiths, Introduction to Elementary Particles,
John Wiley \& Sons Inc. (1987).
\bibitem{Goldman} I.I. Goldman and V.D. Krivchenkov, Problems in Quantum 
Mechanics, Pergamon Press Ltd., 132 (1961).
\bibitem{Ryzhik} I.S. Gradshteyn and I.M. Ryzhik, Table of Integrals, Series,
and Products, Academic Press Inc., 1058 (1980).
\bibitem{PDB00} D. E. Groom {\it et al.}, Review of particle physics, The
European Physical Journal, {\bf C15} (2000). 
\bibitem{SPHINX95} D. V. Vavilov {\it et al.}, Yad. Fiz., {\bf 57},
241 (1994)S; V. Golovkin {\it et al.}, Z. Phys., {\bf C68}, 585 (1995).
\bibitem{SPHINX97} L. G. Landsberg (SPHINX Collaboration), in
Proceedings of the 14th Conference on Particles and Nuclei, edited
by Carl E. Carlson and John J. Domingo, World Scientific, Singapore,
530 (1997).
\bibitem{SPHINX99} L. G. Landsberg (SPHINX Collaboration), in
Proceedings of the 19th International Symposium on Lepton an Photon
Interaction at High Energy, Stanford University, August 9, 1999;
e-print hep-ex/9907025 (1999). 
\bibitem{saphir98} SAPHIR collaboration, M. Bockhorst {\it et al.},
Z. Phys., {\bf C63}, 37 (1994); C. Bennhold {\it et al.},
Nucl. Phys., {\bf A639} (1998).
\bibitem{Gueye00} P. Gu\`eye, in Proceedings of the Hypernuclear physics
with electromagnetic probes (HYPJLAB99) workshop, Hampton, Virginia (2000).
\bibitem{Niculescu98} G. Niculescu {\it et al.}, Phys. Rev. Lett.
{\bf 81}, 1805 (1998).
\bibitem{Mohring03} R. Mohring {\it et al.}, Phys. Rev. {\bf C67} (2003).
\bibitem{Zeidman} B. Zeidman {\it et al.}, CEBAF Proposal PR91-016.
\bibitem{Reinhold98} J. Reinhold {\it et al.}, Nucl. Physics,
{\bf A639}, 197c (1998).
\bibitem{Bennhold99} C. Bennhold and T. Mart, Phys. Rev. C {\bf 61},
012201 (1999).
\bibitem{GW03} P. Gu\`eye \& R. A. Williams, {\it Study of the $X = $(u\sbar)-(uds)
	$S = 0$ pentaquark systems in associated strangeness production around 2~GeV},
	Jefferson Lab Hall C proposal, (2003).
\bibitem{RobCap9498} W. Roberts and S. Capstick, Phys. Rev. D
{\bf 49}, 4570 (1994); and  Phys. Rev. D {\bf 58}, 074011 (1998).
\end{references}
\end{document}